\documentclass[11pt]{article}
\usepackage{amsmath,amssymb,color,graphics,epsfig,cite}

\usepackage{amsfonts}

\newcommand{\be}{\begin{equation}}
\newcommand{\ee}{\end{equation}}
\newcommand{\bea}{\setlength\arraycolsep{2pt} \begin{eqnarray}}
\newcommand{\eea}{\end{eqnarray}}
\newcommand{\nn}{\nonumber}

\def\ft#1#2{{\textstyle{\frac{\scriptstyle #1}{\scriptstyle #2} } }}
\def\fft#1#2{{\frac{#1}{#2}}}

\def\0{{\sst{(0)}}}
\def\1{{\sst{(1)}}}
\def\2{{\sst{(2)}}}
\def\3{{\sst{(3)}}}
\def\4{{\sst{(4)}}}
\def\5{{\sst{(5)}}}
\def\6{{\sst{(6)}}}
\def\7{{\sst{(7)}}}
\def\8{{\sst{(8)}}}
\def\sst#1{{\scriptscriptstyle #1}}

\begin{document}

\begin{center}
{\Large {\bf Thermodynamics of Taub-NUT and Plebanski Solutions}}

\vspace{20pt}

Hai-Shan Liu, H. L\"{u} and Liang Ma

\vspace{10pt}

{\it Center for Joint Quantum Studies and Department of Physics,\\
School of Science, Tianjin University, Tianjin 300350, China }

\vspace{40pt}

\underline{ABSTRACT}
\end{center}

We observe the parallel between the null Killing vector on the horizon and degenerate Killing vectors at both north and south poles in Kerr-Taub-NUT and general Plebanski solutions. This suggests a correspondence between the pairs of the angular momentum/velocity and the NUT charge/potential. We treat the time as a real line such that the Misner strings are physical. We find that the NUT charge spreads along the Misner strings, analogous to that the mass in the Schwarzschild black hole sits at its spacetime singularity. We develop procedures to calculate all the thermodynamic quantities and we find that the results are consistent with the first law (Wald formalism), the Euclidean action and the Smarr relation. We also apply the Wald formalism, the Euclidean action approach, and the (generalized) Komar integration to the electric and magnetic black holes in a class of EMD theories, and also to boosted black strings and Kaluza-Klein monopoles in five dimensions, to gain better understandings of how to deal with the subtleties associated with Dirac and Misner strings.

\vfill{\footnotesize  hsliu.zju@gmail.com \ \ \ mrhonglu@gmail.com\ \ \ liangma@tju.edu.cn }


\thispagestyle{empty}
\pagebreak

\tableofcontents
\addtocontents{toc}{\protect\setcounter{tocdepth}{2}}

\newpage

\section{Introduction}

With Hawking's seminal work on the semiclassical approach to the black hole radiation \cite{Hawking:1974rv}, the black hole dynamics \cite{Bardeen:1973gs} has since been promoted to thermodynamics. It was demonstrated that black hole thermodynamics are consistent with the Euclidean action approach based on the Quantum Statistic Relation \cite{Gibbons:1976ue}. This is different from the usual Euclideanization in Quantum Field Theory, in that the period of the Euclidean time is set by the black hole geometry. This in fact becomes the standard and more convenient way of deriving the Hawking temperature. Iyer and Wald later established that the miraculous first law of black hole thermodynamics had a classical origin as the infinitesimal Hamiltonian \cite{Wald:1993nt,Iyer:1994ys}, and the Bekenstein-Hawking entropy could be interpreted as a Noether charge \cite{Wald:1993nt}. That black holes are thermodynamic systems has now been widely accepted by the theoretical community in General Relativity.

There are however interesting black objects whose thermodynamics are difficult to establish. The most notable one is the Taub-NUT spacetime \cite{Taub:1950ez,Newman:1963yy}. The difficulty arises from the fact that in the Euclidean action approach, the easily-obtained free energy from the on-shell Euclidean action does not parse itself into different thermodynamical variables, except for the temperature and the presumed entropy. Similar issues exist in the Wald formalism, and furthermore the infinitesimal energy is not uniquely defined. In fact, historically, the satisfaction of the first law was established first by obtaining various thermodynamic quantities independently. These include the ADM masses \cite{Arnowitt:1959ah} in asymptotically Minkowski spacetimes. (We shall not consider the cosmological constant in this paper.) In fact, for black holes where the mass can be independently calculated, it can also be easily recognised in the Wald formalism as the closed infinitesimal Hamilitonian, and vice versa.

However, this success meets a great challenge to analyse the Taub-NUT solution, which contains two parameters $(m,n)$. When the NUT parameter $n$ is zero, the solution reduces to the Schwarzschild black hole of mass $m$. However, when $n\ne 0$, the spacetime is not asymptotic to Minkowski spacetime, but only locally flat. Taub-NUT geometry with real-line time have singularities in the form of Misner strings \cite{Misner:1963fr}, even though the spacetime is geodesically complete \cite{Clement:2015cxa}. This means that we do not have an independent way of deriving the mass $M$. Some treated the parameter $m$ as the mass in literature, but this is not universally accepted. Furthermore, there is no independent derivation for the NUT charge $Q_{\rm N}$ and its conjugate potential $\Phi_{\rm N}$. We therefore have a thermodynamic system with five variables, but only the temperature and entropy $(T,S)$ can be well determined, and the rest three variables $(M, Q_{\rm N}, \Phi_{\rm N})$ are unknown. After taking care of the subtleties associated with Misner strings, the free energy from the Euclidean action can be derived, and the consistent mathematical first-order equation relating the integration constants and the horizon data can be established, but these results themselves do not lead to a clear recognition of the three thermodynamic variables.

There have been recently a small surge in the studying of the first law for the Taub-NUT black hole \cite{Hennigar:2019ive,Bordo:2019slw,Bordo:2019tyh,BallonBordo:2019vrn,Durka:2019ajz,
Clement:2019ghi,Wu:2019pzr,Chen:2019uhp,BallonBordo:2020mcs,Awad:2020dhy,Abbasvandi:2021nyv,
Frodden:2021ces,Rodriguez:2021hks,Godazgar:2022jxm,Liu:2022uox,Wu:2022rmx,Awad:2022jgn}. It is after all a simple metric with only two parameters, and once one of the three unknown $(M,Q_{\rm N}, \Phi_{\rm N})$ is determined, the remaining two follow by the satisfaction of the first law. The challenge is to give a well-defined procedure to determine $(M,Q_{\rm N}, \Phi_{\rm N})$ not only for the Taub-NUT metric, but also the more general Kerr-Taub-NUT and even the full Plebanski solution \cite{Plebanski:1975xfb}, which involves not only the Misner but also the Dirac strings \cite{Dirac:1931kp}.

Our results for $(M,Q_{\rm N}, \Phi_{\rm N})$ are different from those in literature. We study the origin of the Misner strings as the degenerate cycles at both north and south poles. The degenerated Killing vectors are given by
\be
\ell_\pm =\partial_\phi \mp 2n \partial_t\,.\label{ellpm0}
\ee
The fact that $\ell_\pm$ both generate $2\pi$ period indicates singularities in the form of Misner strings, since we take time to be a real line. The form of the Killing vectors \eqref{ellpm0} resembles remarkably the null Killing vector in a rotating black hole
\be
\xi = \partial_t + \Omega_+ \partial_\phi\,,\label{nullxi0}
\ee
where $\Omega_+$ is the angular velocity on the horizon for the asymptotically non-rotating Minkowski spacetime. The parallel of these two sets of Killing vectors are striking, with the correspondence
\be
t \leftrightarrow \phi\,,\qquad \Omega_+ \leftrightarrow n\,,\label{duality}
\ee
completed with $r\leftrightarrow u=\cos\theta$, where $(r,\theta)$ are the standard radial and latitudinal coordinates.

It is well known that there exists a symmetry between the radial and latitudinal coordinates and also between time and longitudinal coordinates in the general Plebanski solution \cite{Plebanski:1975xfb}. This symmetry has led people to believe a correspondence between $(m,n)$: the parameter $n$ was regarded as the NUT charge, and it has a physical interpretation as the ``magnetic'' version of the mass. Our correspondence \eqref{duality} differs from this view. Instead, the parameter $n$ should be viewed as NUT potential $\Phi_{\rm N}$. The angular velocity $\Omega_+$ is the thermodynamic conjugate to angular momentum $J$, which can be obtained from the Komar integration associated with the Killing vector $\partial_\phi$, with radially independent integrand. The correspondence then suggests that the NUT charge should be related to the Komar integration associated with the Killing vector $\partial_t$, but now with the latitudinal independent integrand. This provides a well-defined procedure to obtain $(\Phi_{\rm N}, Q_{\rm N})$ not only for the Taub-NUT, but also Kerr-Taub-NUT and the general Plebanski solution, although the results are progressively more complicated. We find that our definition is consistent with the Smarr relation, the Euclidean action and also the Wald formalism, i.e.~the first law. It should be emphasized that in our approach, the NUT potential and its charge $(\Phi_{\rm N}, Q_{\rm N})$ are independently determined, as in the case of $(\Omega_+, J)$, and the satisfaction of the first law arises as a consequence, rather than being used as an input.

The Plebanski solution contains not only the Misner strings but also the Dirac strings, and there are technic subtleties involving these singularities.  In this paper, we begin in section 2 with analysing the magnetically charged black holes in a class of Einstein-Maxwell-Dilaton (EMD) theories and show that the Dirac strings in the evaluation of the Euclidean action, Wald formalism and Komar integration can be dealt in the same way. We then apply the Wald formalism to the Ricci-flat boosted black string and Kaluza-Kaluza-monopole in section 3, where the four-dimensional Dirac strings are lifted to become Misner strings in five dimensions.  After these preliminaries, we finally study our main objects, the Taub-NUT and Kerr-Taub-NUT solutions in section 4. We obtain all the thermodynamic variables first and then verify the first law, the Smarr relation and the free energy from the Euclidean action. We also obtain the thermodynamics for the more general case with asymmetric Misner strings at north and south poles.  In section 5, we continue to successfully analyse the Plebanski solution and obtain the first law that respects the electromagnetic duality. We conclude the paper in section 6. In appendix \ref{app:wald}, we describe the Wald formalism and generalized Komar integration for the EMD theories.

\section{Charged black holes in EMD}
\label{sec:emd}

In this section, we study the electrically or magnetically charged black holes in a general class of EMD theories in four dimensions, with the Lagrangian
\be
{\cal L}=\sqrt{-g} \Big(R - \ft12 (\partial\varphi)^2 - \ft14 e^{a\varphi} F^{\mu\nu} F_{\mu\nu} \Big)\,,\label{emdlag}
\ee
where $F_\2=\fft12 F_{\mu\nu} dx^\mu\wedge dx^\nu =dA_\1$ is the field strength of the Maxwell field $A_\1$.  We use subscript $(k)$ to denote $k$-form fields. The dilaton scalar $\varphi$ has an exponential coupling with the Maxwell kinetic term, with dilaton coupling constant $a$. Such a theory is inspired by string theory and it arises naturally in supergravities with some specific $a$ values. We use this simple model to illustrate how to properly handle Dirac strings associated with the magnetic charges in various thermodynamic calculations.

\subsection{Charged black holes and thermodynamics}

We focus on four dimensions, and the theory can admit either electric or magnetic black holes. The metric takes the same form for both black holes, e.g.~\cite{Duff:1996hp,Cvetic:1996gq, Lu:2013eoa}, given by
\bea
ds^2 &=& -H^{-\fft12{N}} f dt^2  +
H^{\fft12{N}}\big(f^{-1} dr^2 + r^2 (d\theta^2 + \sin^2\theta d\phi^2)\big)\,,\nn\\
H &=& 1 + \fft{q}{r}\,,\qquad f=1- \fft{\mu}{r}\,.
\eea
In this paper, we reserve $(t,r,\theta,\phi)$ to be the time, radial, latitudinal and longitudinal angular coordinates. (We use $\varphi$ to denote the scalar field.) The Maxwell and scalar fields are different, depending on the type of charges, given by
\bea
{\rm electric:}&& A_\1 = \psi_e(r) dt\,,\qquad \psi_e=\fft{\sqrt{Nq(\mu+q)}}{r H}\,,\qquad
\varphi = \ft12 N a\log H\,,\nn\\
{\rm magnetic:}&& A_\1 = \sqrt{Nq(\mu+q)}\, \cos\theta d\phi\,,\qquad \varphi = -\ft12 N a\log H\,.
\label{elec and magn}
\eea
Here the parameter $N$ is a short notation for $N=\fft{4}{a^2+1}$. When $a=0$, corresponding to $N=4$, we have the Reissner-Nordstr\"om (RN) electric or magnetic black holes.  For $a=\sqrt{3}$, $a=1$ or $a=0$, exact solutions of dyonic black holes can also be constructed, but we shall not consider these black holes in this section.

The solution is asymptotic to Minkowski spacetime, and there is an event horizon at $r_+=\mu$. The thermodynamic quantities can be easily obtained, and the mass, temperature and entropy are given by
\be
M=\ft12\mu + \ft14 N q\,,\qquad
T=\fft1{4\pi\mu} (1 + \ft{q}{\mu})^{-\fft12N}\,,\qquad
S=\pi \mu^2 (1 + \ft{q}{\mu})^{\fft12N}\,.\label{MTSforEMD}
\ee
The electric and magnetic charges and their corresponding thermodynamical potentials are
\bea
\hbox{electric}:&& Q_e= \ft14 \sqrt{N q(\mu +q)}\,,\qquad \Phi_e=\sqrt{\fft{Nq}{\mu+q}}\,,\nn\\
\hbox{magnetic}:&& Q_m = \ft14 \sqrt{N q(\mu + q)}\,,\qquad \Phi_m=\sqrt{\fft{Nq}{\mu+q}}\,.
\eea
It is then straightforward and simple to verify that the first law of black hole thermodynamics holds for both the electric and magnetic black holes, namely
\bea
\hbox{electric}:\qquad \delta M=T\delta S + \Phi_e \delta Q_e\,;\qquad \hbox{magnetic}:\qquad \delta M=T\delta S + \Phi_m \delta Q_m\,.\label{emdfl}
\eea

In the above discussion, even though we had to appeal to the electromagnetic duality to derive the magnetic potential, all the thermodynamical quantities can be independently obtained. The resulting first law therefore appears to be miraculous. In the next, we shall review both the Euclidean path integral approach and the Wald formalism that underly the first law. We shall focus on the technical subtleties involving the magnetic monopoles.

\subsection{Euclidean action}

Based on the Quantum Statistic Relation, one can use a path integral approach to the black hole thermodynamics by calculating the on-shell Euclidean action \cite{Gibbons:1976ue}
\be
I = \fft{1}{16\pi} \int_{\cal M} d^4 x  {\cal L} + \fft{1}{8\pi} \int_{\partial {\cal M}} d^3x\,\sqrt{-h} K = \beta F\,,
\ee
which gives rise to the free energy $F$.  (There should be no confusion between this free energy symbol and Maxwell field strength $F_\2$.) The corresponding thermodynamic ensembles for the electric and magnetic cases are very different, with the former corresponding to the Gibbs free energy whilst the latter to the Helmholtz free energy:
\be
{\rm electric:}\qquad  F_{\rm G} = M - T S - \Phi_e Q_e\,,\qquad
{\rm magnetic:}\qquad  F_{\rm H} = M - T S\,.
\ee
In the above discussion, we assume that the variation principle for the Maxwell field obeys the standard Dirichlet boundary condition. We can alter it to the Neumann boundary condition by introducing a surface term \cite{Gibbons:1976ue,Hawking:1995ap,Reall:2001ag,goto}
\be
\fft{1}{16\pi} \int_{\partial {\cal M}} d\Sigma_\mu\left( e^{a\varphi} F^{\mu\nu} A_\nu\right).\label{emboundary}
\ee
This is equivalent to introducing a total derivative bulk term
\be
\fft{1}{16\pi} \int_{\cal M} d^4x\,\sqrt{-g} \nabla_\mu\left(e^{a\varphi} F^{\mu\nu} A_\nu\right).\label{emtotd}
\ee
From the first-order formalism \cite{Liu:2019smx}, adding this term effectively performs the electromagnetic duality and therefore we have
\be
{\rm electric:}\qquad  F_{\rm H} = M - T S\,,\qquad
{\rm magnetic:}\qquad  F_{\rm G} = M - T S - \Phi_m Q_m\,.
\ee
However, in the case of the magnetic black hole, there is a subtlety if we evaluate the on-shell action using the boundary term \eqref{emboundary} instead of the bulk total derivative term \eqref{emtotd}. The bulk total derivative term \eqref{emtotd} gives
\bea
\fft{1}{16\pi} \int_{\cal M} d^4x\,\sqrt{-g} \nabla_\mu\left(e^{a\varphi} F^{\mu\nu} A_\nu\right)=\frac{N q}{4}\int dt= \Phi_m Q_m \int dt\,.\label{emdbulk}
\eea
This bulk integration is unambiguous since it is gauge invariant. The derivative in the integrand can only act on the Maxwell field for the on-shell fields.

The integration on the surface, which should yield the same result, is much subtler. For magnetic charges, both $A_\1$ and $F_\2$ have no components in the radial direction. One would thus na\"ively conclude that the surface term vanishes. (We shall comment on this at the end of this subsection; a plot twist shows that it is not so na\"ive after all.) To resolve this paradox, one notes that bare gauge potential $A_\1$ appears in the expression \eqref{emboundary}, indicating that it is not manifestly gauge invariant. For the monopole \eqref{elec and magn}, two symmetric Dirac strings \cite{Dirac:1931kp} are located at north and south poles $\theta=(0,\pi)$ respectively, where the integrand \eqref{emboundary} is singular
\bea
-\frac{1}{2}e^{a\varphi} F^{\mu\nu} A_\nu&&=\frac{N q  (\mu +q)}{2 r^4}e^{a \varphi } H^{-N}\frac{\cos{\theta}}{\sin{\theta}}\delta_\theta^\mu\,.
\eea
Thus in addition to the integration over the surface normal to the radial direction that vanishes in this example, we also need to integrate over the infinitesimally-thin tubes $T_N$ and $T_S$ at $\theta=0,\pi$ that cutout the Dirac strings. (See the illustration of the fourth graph of Fig.~\ref{plots} in appendix \ref{app:wald}.) For the surface normal to constant $\theta$, we have
\bea
d\Sigma_\mu=dtdrd\varphi\sqrt{-h^\theta}n_\mu^\theta\,,\qquad \sqrt{-h^\theta}=rH^{\frac{N}{4}}\sin{\theta}\,,\qquad
n^\theta=-rH^{\frac{N}{4}}d\theta\,.
\eea
We can compute \eqref{emboundary} by the integration over the two tubes,
\bea
&&\fft{1}{16\pi} \int dt\int_{0}^{2\pi}d\phi\Bigg[\int_{r_h}^{+\infty} dr\sqrt{-h^\theta}n_\mu^\theta\left(e^{a\varphi} F^{\mu\nu} A_\nu(\theta=0)\right)\cr
&&+\int^{r_h}_{+\infty} dr\sqrt{-h^\theta}n_\mu^\theta\left(e^{a\varphi} F^{\mu\nu} A_\nu(\theta=\pi)\right)\Bigg]=\frac{N q}{4}\int dt\,.
\eea
This yields the exact same result of the bulk integration \eqref{emdbulk}. The first and second terms in the integrand are associated with $T_N$ and $T_S$ respectively. They contribute equally under our gauge choice where the Dirac strings are symmetric. We can make gauge choices such that the Dirac strings only appear at the north pole or at the south pole, as indicated in the second and third graphs in Fig.~\ref{plots}. In these cases we only need to integrate over $T_N$ or $T_S$ only, but with twice the value, so that the final answer remains the same.  This approach gives the same answer as the method of double covering space adopted in \cite{Wu:1975es,Yang:1996cla}.

This subtlety turns out to be also relevant to understand the Wald approach to these magnetic black holes, since in essence, the Wald formalism also turns the bulk integration to surface integration. The above discussion also suggests a method for the surface integration that sometimes are much simpler and unambiguous: turning the surface integration into a bulk integration that is gauge invariant.

Before continuing, it is important to clarify that in the approach above and through out this paper, we treat the string singularities as being real and physical, and we cage them inside the infinitesimal tubes $T_N$ and $T_S$ in our evaluation of the boundary term. In this approach, the boundary term gives rise the same result as the corresponding bulk term, by the virtual of the Stokes theorem. The boundary term has the same effect of performing the electromagnetic duality transformation. The price to pay is that the free energy from the Euclidean action is not invariant under the electromagnetic duality. Indeed, the electromagnetic symmetry is broken, since the magnetic charge produces the string singularity, while the electric charge does not, and the electromagnetic duality transformation cannot restore this symmetry. An alternative approach is to observe that in Einstein-Maxwell gravity and EMD, all the fields are neutral. The Dirac string that appears in the surface term is really an artificial introduction. One can define the manifold with double patches so that the Dirac strings are completely absent in the manifold \cite{pang}. In this case, the vanishing result of the boundary term for magnetic monopoles, which can be obtained from the na\"ive application mentioned earlier, is not only actually correct, but also desirable.  In this approach, the free energy for both electric and magnetic black holes takes the same form $F_{\rm H} = M - T S$, which is manifestly invariant under the electromagnetic duality \cite{pang}. However it is not clear to us how this approach is consistent with the Stokes theorem. Nevertheless, since the two approaches differ by a thermodynamic Legendre transformation, it does not affect what we shall address in this paper.

\subsection{Wald formalism}

In 1993, Iyer and Wald developed the covariant phase space method \cite{Wald:1993nt, Iyer:1994ys} to point out that the first law of black hole thermodynamics can be encoded in the infinitesimal Hamiltonian. The Wald procedure for the EMD theory \eqref{emdlag} is described in appendix \ref{app:wald}. The Wald approach to the first law for the electrically charged black holes is also given in the appendix.  The magnetic case is subtler and we study it here. As we discuss in the appendix, the crucial step of the Wald formalism is to calculate the 2-form fields ${\mathbf Q}[\xi]$ and $i_\xi {\mathbf \Theta}$ with $\xi=\partial_t$. (Note that the boldface letters in this paper are also form fields.) For the magnetic solution, we have
\be
\mathbf{Q}[\xi]=r^2\left(f'-\frac{NfH'}{2H}\right)\,\Omega_\2\,,\qquad
\delta \mathbf{Q}[\xi] - i_\xi\mathbf{\Theta} = V \Omega_\2 + U dr\wedge \cos\theta d\phi\,,
\label{emdmuv}
\ee
where $\Omega_\2 = \sin\theta d\theta\wedge d\phi$ volume 2-form of the round unit $S^2$, and
\be
U=-\fft{16  Q_m \delta Q_m}{r^2 H^2}\,,\qquad V= \fft{r}{2H}\Big(N r f' \delta H-2 N r f \delta H'-\left(N r H'+4 H\right)\delta f\Big)\,.
\ee
As in the electric case discussed in the appendix, $\mathbf{Q}[\xi]$ is not close, i.e. $d\mathbf{Q}\ne 0$, but $\delta \mathbf{Q} - i_\xi\mathbf{\Theta}$ is, namely
\be
d(\delta \mathbf{Q}[\xi] - i_\xi\mathbf{\Theta}) = (V' + U) dr\wedge \Omega_\2 =0\,.\label{emdmdqth}
\ee
Although we can write the $\delta {\cal H}$ formally as
\be
0=\delta {\cal H} =\fft{1}{16\pi} \int_{\rm vol}  d(\delta {\mathbf Q}[\xi]
- i_\xi\mathbf{\Theta}) =\fft{1}{16\pi} \int_{\Sigma} \Big(V \Omega_\2 + U dr\wedge \cos\theta d\phi\Big)\,,
\ee
the surface symbol ``$\Sigma$'' no longer denotes only the 2-spheres of the asymptotic infinity and the horizon Cauchy surface. The second term associated with $U$ function is singular at both north $\theta=0$ and south $\theta=\pi$ poles in the form of strings, extending from the horizon to asymptotic infinity, as in the case of the Dirac strings.  The integration of this term is mathematically the same as the Maxwell surface term for the magnetic charge in the Euclidean action discussed earlier. The full integration should be split into four parts: the integrations over $\Omega_\2$ both on the horizon and at asymptotic infinity; together with integrations over the tubes at the north and south poles $(\delta{\cal H}_{T_N}, \delta{\cal H}_{T_S})$ that remove the Dirac strings. For the first two parts, we have
\be
\delta{\cal H}_{S_2}^+= -\ft14 V(r_+) = -T \delta S\,,\qquad \delta{\cal H}_{S_2}^\infty=\delta M\,.
\ee
For the latter two parts, we have
\be
\delta{\cal H}^{\theta=\pi}_{T_S}=\delta{\cal H}^{\theta=0}_{T_N} = \fft{1}{16\pi} \int_{r_+}^{\infty}U(r) dr \int_{0}^{2\pi}d\varphi=-\ft12\Phi_m\delta Q_m\,.
\ee
Combining these four parts, we obtain the relation for the infinitesimal Hamiltonian
\be
\delta{\cal H}_{\Sigma} = \delta{\cal H}^{\infty}_{S_2}+\delta{\cal H}^{+}_{S_1}+\delta{\cal H}^{\theta=0}_{T_N}+\delta{\cal H}^{\theta=\pi}_{T_S}=0\,.
\ee
This leads to the first law given in \eqref{emdfl}.

As in the case of its Euclidean action calculation, we can also have a simpler approach starting from \eqref{emdmdqth}, which implies that we can have a different $(\delta \mathbf{Q}[\xi] - i_\xi\mathbf{\Theta})$, given by
\be
\delta \mathbf{Q}[\xi] - i_\xi\mathbf{\Theta} = (V + {\cal U}) \Omega_\2\,,\qquad
{\cal U} = \int^r U(r') dr' = \fft{16 Q_m \delta Q_m}{r H}\,.\label{emdmnuv}
\ee
This expression differs from the one in \eqref{emdmuv} by a total derivative, i.e.~a closed 2-form, but now the Dirac string singularity is absent. The first law \eqref{emdfl} is then a simple consequence of
\be
V(r_+) + {\cal U}(r_+) = V(\infty) + {\cal U}(\infty)\,.
\ee
In fact, there is a covariant approach to add the needed total derivative to $(\delta \mathbf{Q} - i_\xi\mathbf{\Theta})$ in \eqref{emdmuv} to become \eqref{emdmnuv}. As in the case of Euclidean action, this is effectively equivalent to performing the electromagnetic duality. As in \cite{Lu:2013ura,Ma:2022nwq}, from the equation of motion of the Maxwell field
\bea
d\big(e^{a\varphi}{{*F}_\2}\big)=0\,,
\eea
we can define a scalar field $\Psi$
\bea
d\Psi=e^{a\varphi}i_\xi {*F}_\2\,, \qquad \Rightarrow\qquad \Psi=\frac{4Q_m}{rH}.
\eea
We can then insert the improved total derivative term $-d(\Psi\delta A_{\1})$, that is closed but not exact, into the integrand of $\delta {\cal H}$:
\bea
\delta{\cal H}_{\Sigma}=\frac{1}{16\pi}\int_{\Sigma}( \delta\mathbf{Q}[\xi]-i_{\xi}\mathbf{\Theta}-d(\Psi\delta A_{\1})) = \frac{\int \Omega_\2}{16\pi}\,
(V+ {\cal U})\Bigg|_{r_+}^\infty\,.
\eea
This simple example helps us greatly on how to extract the charges from the closed form fields
in general type-D metrics that are cohomogeneity two and depend on both the radial and latitudinal angular coordinates.

\subsection{Generalized Komar integration and the Smarr relation}

When matter fields are involved, the 2-form $\mathbf Q[\xi]$ in the Wald formalism is not closed. In order to obtain the Smarr relation that relates the asymptotic charges to the horizon data, we need to consider the generalized Komar integration over the closed form field $\widetilde {\mathbf Q}[\xi]$. For the EMD theory, it is given by \eqref{emdtildeQ} in the appendix. As we see in the appendix, the $\widetilde {\mathbf Q}[\xi]$-form is radially independent for the electric black hole.  For the magnetic black hole we study in this section, it is trickier, given by
\be
\widetilde {\mathbf Q}[\xi] = V\Omega_\2 + U dr\wedge \cos\theta d\theta\,,
\ee
with
\be
V=r f' - \fft{N r^2 f H'}{2H}\,,\qquad U=-\fft{Nq(\mu+q)}{2r^2 H^{\fft12N}} e^{a\varphi}\,.
\ee
The closure of $d \widetilde {\mathbf Q}[\xi]=0$ is satisfied by the on-shell identity $V' + U =0$. Thus we can deal with the generalized Komar integration the same way we did with the closed Wald form $(\delta \mathbf{Q}[\xi] - i_\xi\mathbf{\Theta})$ earlier, namely
\be
0=\fft{1}{8\pi} \int_{\Sigma} {\mathbf Q}[\xi]=M - 2TS - \Phi_m Q_m\,,
\ee
which is precisely the Smarr relation. It should be commented here that the reason why we do not redefine the mass to $(M - \Phi_m Q_m)$ is due to the fact that the term $\Phi_m Q_m$ is contributed by the the Dirac string, which is not universal and not experienced by neutral particles.

As we have seen in the appendix, $\widetilde {\mathbf Q}[\xi]$ is radially independent for the electric black holes. To obtain the radially independent charge for the magnetic case is more involved. This subtlety can be avoided entirely by performing the electromagnetic duality, in which case, the generalized Komar form becomes
\be
\widetilde {\mathbf Q}[\xi] = -{*d}\xi-\ft{1}{2}e^{-a\varphi}{*\widetilde F}_{\2}\left(i_\xi B_{\1}\right)+\ft{1}{2}e^{-a\varphi}\left(i_\xi {*\widetilde F}_{\2}\right)\wedge B_\1\,,\label{emdtildeQmag}
\ee
where $\widetilde F_\2$ and $B_\1$ are defined by
\be
e^{a\varphi} {*F}_\2 = \widetilde F_\2 = dB_\1\,.
\ee
For the magnetic solution, we now have
\be
\widetilde {\mathbf Q}[\xi] =\Big(V(r) + \int_\infty^r U(r') dr'\Big) \Omega_\2 = 2M\Omega_\2\,,
\ee
which is indeed radially invariant.

Note that $\widetilde {\mathbf Q}[\xi]$'s in \eqref{emdtildeQmag} and \eqref{emdtildeQ} are both closed; they differ by a total derivative $d{\mathbf \Lambda}_\1$, with ${\mathbf \Lambda}_\1 =(i_\xi B_\1) A_\1$. However, neither expression is invariant under the electromagnetic duality
\be
A_\1 \rightarrow B_\1\,,\qquad B_\1\rightarrow -A_\1\,,\qquad \varphi\rightarrow -\varphi\,.
\ee
The invariant expression can be obtained by taking ${\mathbf \Lambda}_\1 =\fft12 (i_\xi B_\1) A_\1$, which yields
\be
\widetilde {\mathbf Q}[\xi] = -{*d}\xi-\ft{1}{2}e^{a\varphi}{* F}_{\2}\left(i_\xi A_{\1}\right) -\ft{1}{2}e^{-a\varphi}{*\widetilde F}_{\2}\left(i_\xi B_{\1}\right)\,.\label{emdtildeQemsym}
\ee

To summarize, the basic mathematics for evaluating the first law from the Wald formalism and the Smarr relation from the generalized Komar integration is the same Stokes theorem and hence the technique of dealing the string singularities is also the same:
\bea
&&\hbox{First law}:\qquad d(\delta \mathbf{Q}[\xi] - i_\xi\mathbf{\Theta})=0\,,\qquad
\rightarrow \qquad \fft{1}{16\pi} \int_\Sigma (\delta \mathbf{Q}[\xi] - i_\xi\mathbf{\Theta})=0\,,\nn\\
&&\hbox{Smarr relation}:\qquad d \widetilde {\mathbf Q}[\xi]=0\,,\qquad
\rightarrow\qquad \fft{1}{8\pi} \int_{\Sigma} \delta \widetilde {\mathbf Q}[\xi]=0\,.
\eea
Note that when $a=0$, the theory effectively reduces to Einstein-Maxwell gravity, since the dilaton decouples from the solution. The corresponding results will be used again for the studying of the general Plebanski solution in section \ref{sec:pleb}.

Before finishing this section, we would like to comment further on the Komar and generalized Komar integrations. When matter is involved, we must consider the generalized Komar integration in order to obtain the Smarr relation. However, it is not uncommon that one can get the correct mass by evaluating the Komar integration at the infinity. This can indeed be done when the topology of the black hole is simple such that the hypersurface $\Sigma$ includes only $S_1$ on the horizon and $S_2$ at the infinity, as illustrated in the first graph in Fig.~\ref{plots} in the appendix. In this case, since the matter contribution to ${\mathbf Q}$ tends to fall off faster, one can obtain the mass by simply integrating the Komar form over $S_2$ at infinity, without having to know the full generalized Komar form. However, when there are Misner strings on the bulk, these configurations can also contribute to the mass. One can thus no longer obtain the correct mass by only evaluating the Komar integration at infinity.

\section{Boosted black string vs.~Kaluza-Klein monopole}
\label{sec:d=5}

For the EMD theory discussed in the previous section, when the dilaton coupling constant $a=\sqrt3$, the theory can be obtained from the $S^1$ reduction of five-dimensional pure gravity. In this case, the Maxwell field is the Kaluza-Klein vector and the dilaton is the breathing mode of the internal circle.  Consequently, all the four-dimensional solutions can be lifted to become the Ricci-flat solutions in five dimensions. In particular, the electrically and magnetically charged black holes become the boosted black string and Kaluza-Klein monopole respectively. In this section, we would like to study their first laws in five dimensions and recover the ones obtained in $D=4$.

\subsection{Boosted black string}
\label{sec:boost}

The Ricci-flat metric in five dimensions is
\bea
ds^2_5 &=& - H(r)^{-1} f(r) dt^2 + H(r) (dx -w(r) dt)^2 + \fft{dr^2}{f(r)} +
r^2 (d\theta^2 + \sin^2\theta\, d\phi^2)\,,\nn\\
H&=& 1 + \fft{q}{r}\,,\qquad f=1 - \fft{\mu}{r}\,,\qquad w=\fft{\sqrt{q(\mu+q)}}{rH}\,.
\eea
The solution can be also obtained from the static black string $(q=0)$ by a Lorentz boost along the string direction $x$ with $\sinh\delta = q/\mu$. We have chosen the coordinate gauge such that string is non-moving asymptotically, i.e.~$w(\infty)=0$. The horizon is located at $r_+=\mu$, with the null Killing vector
\be
\xi = \partial_t + \Phi_v \partial_x\,,\qquad \Phi_v=\sqrt{\fft{q}{\mu+q}}\,.
\ee
In other words, the horizon, described by $x$ and $(\theta,\phi)$ coordinates, is $\mathbb{R}\times S^2$, and it is moving along the $\mathbb{R}$ direction, with velocity $\Phi_v$. The surface gravity $\kappa$ and corresponding temperature and entropy can be obtained from the standard method, given by
\be
\kappa^2 = -\fft{g^{\mu\nu}\partial_\mu \xi^2\partial_\nu \xi^2}{4\xi^2}\,,\qquad
T=\fft{\kappa}{2\pi}=\frac{1}{4 \pi\sqrt{\mu(\mu+q)}}\,,\qquad S=\pi  \mu ^{3/2} \sqrt{\mu +q}\,.\label{surfgrav}
\ee
The linear momentum $Q_v$ can be obtained from the Komar integration associated with the Killing vector $\partial_x$, given by
\be
Q_v=\frac{1}{4} \sqrt{q (\mu +q)}\,.
\ee
We can then obtain the first law
\be
\delta M = T \delta S + \Phi_v \delta Q_v\,,\qquad M = \ft12 \mu + \ft14 q\,.\label{bsfirstlaw}
\ee
Note that we obtain the mass above by the requirement of the first law; it is not obtained from the Komar integration associated with the Killing vector $\partial_t$, which turns out to be proportional to $(\mu + q)$.

We now would like to derive the first law \eqref{bsfirstlaw} from the Wald formalism which we describe in the appendix. We are now dealing with $D=5$ and hence ${\mathbf Q}[\xi]$ and $i_\xi {\mathbf \Theta}$ are 3-forms. It is convenient to define a close 3-form
\be
\Omega_\3 = \sin\theta d\theta\wedge d\phi \wedge (dx - \Phi_v dt)\,.
\ee
We then have
\bea
{\mathbf Q}[\partial_t] &=& r^2 \left(f'-H^2 \omega \omega'-f \fft{H'}{H}\right)\Omega_\3=(\mu+q)\Omega_\3\,,\nn\\
{\mathbf Q}[\partial_x] &=& H^2 r^2 \omega'\, \Omega_\3= -4 Q_v \Omega_\3\,.\label{boostq}
\eea
This leads to
\be
{\mathbf Q}[\xi] = (\mu+q - 4 \Phi_v Q_v) \Omega_\3
=\mu \Omega_\3\,.
\ee
Note that in pure gravity, the $\mathbf Q$ in the Wald formalism is the same as the Komar form and it is closed. We now proceed and find
\bea
i_\xi {\mathbf \Theta} &=& -\fft{r}{2H^2}\Big(2  H^4 r \omega' \delta \omega + r \left(H f'-2 f H'\right)\delta H-2 H^2 r \delta f'+ H \left(r H'-4 H\right)\delta f \Big)\Omega_\3\nn\\
&=&-\delta \mu \Omega_\3\,.\label{boostitheta}
\eea
Thus the combination yields
\be
\delta {\mathbf Q}[\xi] - i_\xi {\mathbf \Theta} = 4\delta(M - \Phi_v Q_v)\,\Omega_\3\,.
\ee
This shows that this combination is indeed a close 3-form; therefore, it is radially conserved, i.e.
\be
\delta {\cal H} = \fft{1}{16\pi} \int_\Sigma (\delta {\mathbf Q}[\xi] - i_\xi {\mathbf \Theta})=\delta {\cal H}_\infty -
\delta {\cal H}_+=0\,.
\ee
In particular, we have
\be
\delta {\cal H}_{\infty} = \delta {\cal H}_+ = \delta(M - \Phi_v Q_v)={\rm constant}\,.\label{deltainfinity}
\ee
It appears that the dynamical law is a trivial identity. This is because we have simply substituted the solution into the form fields, and the radially conserved quantity is necessarily expressed as some integration constants of the solution. In order to read off the horizon data, we should evaluate the 3-form $(\delta {\mathbf Q}[\xi] - i_\xi {\mathbf \Theta})$ in terms of the abstract functions $(H,f,w)$.  On the horizon, we can impose $f(r_+)=0$ and $w(r_+)=\Phi_v$. Furthermore, for any function $\chi(r)$ that is regular on the horizon, its variation on the horizon
is
\be
\delta \chi\Big|_{r=r_+} = \delta (\chi(r_+)) - \chi'(r_+) \delta r_+ \,.\label{varychi}
\ee
With these, it follows from \eqref{boostq} and \eqref{boostitheta} that we obtain
\be
\delta {\cal H}_+ = T \delta S - Q_v\delta \Phi_v\,.\label{deltahorizon}
\ee
The first law \eqref{bsfirstlaw} is then the consequence of \eqref{deltainfinity} and \eqref{deltahorizon}.  Thus we learn that substituting an exact solution into the form field $(\delta {\mathbf Q}[\xi] - i_\xi {\mathbf \Theta})$ will simply yield a radially conserved quantity of integration constants that are typically recognisable as asymptotic data. The horizon data should be extracted from the near-horizon geometry.

We verify that the free energy associated with the Euclidean action is given by
\be
F_{\rm G}=M - T S - \Phi_v Q_v\,.\label{bsfree}
\ee
We therefore recover the thermodynamics of the corresponding EMD black hole, without the Maxwell boundary term \eqref{emboundary}. This purely geometric example will help us to study more complicated rotating metrics discussed later.

\subsection{Kaluza-Klein monopole}
\label{sec:kkmono}

The Ricci-flat Kaluza-Klein monopole is obtained by lifting the $a=\sqrt{3}$ magnetically-charged black hole to five dimensions. The five-dimensional metric is
\bea
d\hat{s}^2_5 &=&-fdt^2+H\left(\frac{dr^2}{f}+r^2(d\theta^2 + \sin^2\theta d\phi^2)\right)+H^{-1}
\left(d\psi+P\cos{\theta}d\phi\right)^2\,,\nn\\
H &=& 1 + \fft{p}{r}\,,\qquad f=1 - \fft{\mu}{r}\,,\qquad P=\sqrt{p(p+\mu)}\,.
\eea
The metric is asymptotic to a constant $U(1)$ bundle over the four-dimensional Minkowski spacetime. For the solution to be absent from a string singularity, we must require that the fibre coordinate $\psi$ have a period of
\be
\Delta \psi = 4\pi P\,.\label{deltapsi}
\ee
However, we would like first to interpret the solution as the lift from the magnetic black hole from $D=4$, in which case, the $\Delta \psi$ is fixed, independent of the integration constants that will be interpreted as thermodynamic variables. In fact, the period $\Delta\psi$ will be absorbed into the four-dimensional Newton's constant upon dimensional reduction and hence does not involve in the black hole thermodynamic first law in $D=4$. For now, we shall
consider the case with fixed $\Delta \psi$ and without loss of generality, we set $\Delta \psi=1$. This implies that the metric is singular, and the singularity can be best described as Misner strings attached to the north and south poles, associated with the singularity of the connection $P \cos\theta d\phi$ in the $U(1)$ fibre $\psi$.

The metric has a horizon located at $r_+=\mu$ and the temperature and entropy is given by
\be
T= \fft{1}{4 \pi \sqrt{\mu(\mu+p)}}\,,\qquad S=\pi \mu^{\fft32} \sqrt{\mu + p}\,.
\ee
However, it is not easy to complete the first law based on this information, without knowing that the solution is related to the four-dimensional magnetic black hole. Making use of the fact that $P$ describes the charge, one can derived the mass and first law.  We now examine the dynamics using the Wald formalism. We find that
\be
{\mathbf Q}[\xi] = r^2 f' \Omega_\3 = \mu \Omega_\3\,,\qquad \Omega_\3 = \sin\theta d\theta\wedge d\phi \wedge d\psi\,,\qquad \xi=\partial_t\,.\label{monopoleQ}
\ee
This Komar 3-form is proportional to $\mu$, which however is not expected to be the mass. The closed 3-form associated infinitesimal variation of the Hamiltonian is
\bea
&&\delta {\mathbf Q}[\xi] - i_\xi {\mathbf \Theta} = V(r) \Omega_\3 + U(r) dr\wedge \cos\theta d\phi\wedge d\psi\,,\nn\\
&&V = -\fft{r}{2H} \Big(4 \delta f H-r \left(\delta H f'-2 f \delta H'-\delta f H'\right)\Big)\,,\qquad U = -\frac{P\delta P}{H^2 r^2}
\eea
It is clear that $d(\delta {\mathbf Q}[\xi] - i_\xi {\mathbf \Theta})=0$, since we have the on-shell identity $V' + U=0$. This implies that we must have
\be
\delta {\cal H} = \fft{1}{16\pi} \int_{\Sigma} \Big(\delta {\mathbf Q}[\xi] - i_\xi {\mathbf \Theta}\Big)=0\,.
\ee
While it is straightforward to integrate the $V$ term, the integration of the $U$ term is subtle. It is analogous to the situation with the magnetic charge described in section 2.  Similarly, we can adopt two approaches: one is to integrate over tubes $T_N$ and $T_S$ connecting to the north $(\theta=0)$ and south $(\theta=\pi)$ poles. Alternatively we can introduce a close but not exact form such that the $U$ term becomes well defined, namely
\bea
(\delta {\mathbf Q}[\xi] - i_\xi {\mathbf \Theta}) &\rightarrow& (\delta {\mathbf Q}[\xi] - i_\xi {\mathbf \Theta})  - d({\cal U} \cos\theta d\phi\wedge d\psi)=
(V + {\cal U}) \Omega_\3\,,\nn\\
&& {\cal U} = \int^{r}_\infty dr U = \fft{P\delta P}{rH}\,,
\eea
It is then easy to verify that
\be
V\Big|_{\infty} = 4 \delta (\ft12\mu + \ft14p)\equiv 4\delta M \,,\qquad V\Big|_{+} = 4 T \delta S\,,\qquad
{\cal U}\Big|_{\infty}=0\,,\qquad {\cal U}\Big|_{+}=4 \Phi_p \delta Q_p\,,
\ee
where $Q_p=P/4$ and $\Phi_p=P/\mu$. This leads to first law of the Ricci-flat Kaluza-Klein monopole
\be
\delta M= T \delta S + \Phi_p \delta Q_p\,.
\ee
It takes the identical form as the one of the magnetic black hole in four dimensions. This Kaluza-Klein monopole is a particularly interesting example in that the subtlety of the Dirac strings is lifted to become that of Misner strings. Indeed, in the Wald formalism, the Dirac strings and Misner strings have the same mathematic form, and hence can be dealt with in the same way. The free energy from the Euclidean action can also be calculated straightforwardly, given by
\be
F_{\rm H} = M - T S\,.\label{kkfree}
\ee
It is worth comparing to the free energy of the boosted string \eqref{bsfree}; now there is no Legendre transformation associated with ($Q_p, \Phi_p$).

The closure of Komar 3-form ${\mathbf Q}[\xi]$ in \eqref{monopoleQ} implies that
\be
\mu = 4 T S\,.
\ee
This is equivalent to the Smarr relation $M=2 T S + \Phi_p Q_p$, since we have $\Phi_p Q_p=q /4$.  This illustrate the danger of reading off the mass from the Komar integration when the metric is not asymptotic to the Minkowski spacetime.

Finally we would like to comment the case when we do take the period of $\psi$ to be \eqref{deltapsi}. We can redefine the coordinate $\psi = P \widetilde \psi$ and the metric becomes
\be
d\hat{s}^2_5 =-fdt^2+H\left(\frac{dr^2}{f}+r^2(d\theta^2 + \sin^2\theta d\phi^2)\right)+H^{-1}
P^2\left(d\widetilde \psi+\cos{\theta}d\phi\right)^2.\nn\\
\ee
Now the solution is smooth without the Misner singularity, with the level surfaces being the smooth squashed 3-spheres.  The variable $P$ describes the radius of the $U(1)$ fibre of the three sphere. The temperature remains the same, but entropy and mass are modified
\be
T= \fft{1}{4 \pi \sqrt{\mu(\mu+p)}}\,,\qquad S=4\pi^2 \mu^{\fft32} (\mu + p)\sqrt{p}\,,\qquad
M=\pi (2\mu+p)\sqrt{p(\mu+p)}\,.
\ee
We now have a new pair of thermodynamic variables, the circumference of the $U(1)$ fibre and the conjugate tension force $F_{\rm T}$
\be
L=4 \pi \sqrt{p(\mu+p)}\,,\qquad F_{\rm T}=\fft14 (\mu + 2p)\,.
\ee
The first law of the smooth Kaluza-Klein monopole without Misner strings now becomes
\be
\delta M = T \delta S + F_{\rm T} \delta L\,.
\ee
Note that the extensive quantities $(M,S)$ are multiplied by an $L$ factor, compared to those in the Kaluza-Klein monopole with Misner strings. The free energy from the Euclidean action is again the form of \eqref{kkfree}.

Thus we see that there are two globally distinct Ricci-flat metrics, even though they are the same locally. One is smooth without Misner string and it should not be called as the monopole. The proper Kaluza-Klein monopole has Misner strings and its monopole charge is really the doppelganger of the magnetic charge of the corresponding four-dimensional black hole.

\section{Taub-NUT and Kerr-Taub-NUT}
\label{sec:nutknut}

We are now in the position to study the thermodynamic first law of the four-dimensional Taub-NUT metric \cite{Taub:1950ez,Newman:1963yy}, and its Plebanski generalization \cite{Plebanski:1975xfb}. In this section, we focus on the Ricci-flat metrics. We shall apply the Wald formalism to the Taub-NUT and Kerr-Taub-NUT. As was discussed earlier, the Wald-formalism establishes a first-order differential relation among the integration constants, but it does not always give a clear physical interpretation of these constants.
We shall therefore use a variety of tools and techniques outlined in the previous sections.
Our reading of the first law differs from those in literature. We find that the more general Plebanski solutions appear to confirm our interpretation.

\subsection{Taub-NUT}
\label{sec:nut}

The Ricc-flat Taub-NUT solution is given by
\be
ds_4^2=-f\left(dt+2n \cos\theta d\phi\right)^2+\frac{dr^2}{f}+ \left(r^2+n^2\right)\left(d\theta^2+\sin^2\theta d\phi^2\right),\quad f=\frac{r^2-2m r-n^2}{r^2+n^2}\,.\label{Taub-NUT}
\ee
The solution contains two integration constants $(m,n)$. The parameter $m$ can be easily recognized as the condensation of the massless graviton. The parameter $n$, typically referred to as the NUT charge, has an obscure physical meaning and to find the right form of the first law may shed light on its meaning.  Our finding indicates that $n$ is not the NUT charge, but the thermodynamic potential conjugate to the NUT charge. For this reason, we shall refer $n$ to the NUT parameter.

The solution contains only two parameters. If one can determine the mass, the first law is then not difficult to decode. However, the Taub-NUT metric is not asymptotic to Minkowski spacetime at large $r$, but some locally asymptotic flat spacetime.  This makes the definition of the mass difficult. The parameter $m$, associated with the graviton condensation, may be referred to as the gravitational mass in \cite{Lu:2014maa}, and was indeed treated as the mass of the Taub-NUT solution in some works in literature, e.g.~\cite{Hennigar:2019ive,Bordo:2019slw,
Frodden:2021ces,Bordo:2019tyh,BallonBordo:2020mcs,Abbasvandi:2021nyv}. (In particular, this mass was also independently calculated via the conformal completion method \cite{Bordo:2019tyh}.) However, we disagree with this interpretation. It is important to note that the solution has a bizarre property that it describes a black object with event horizon for all real values of $m$.  In other words, for nonvanishing $n$, there is a horizon for all $m\in (-\infty,\infty)$. To have no lower bound of mass is not a satisfactory concept for black hole objects, even though this does not necessarily violate the positive energy theorem, which typically requires asymptotically Minkowski spacetime.

In our approach, the temperature and entropy are calculated in a traditional sense. (New approach was recently proposed in \cite{Godazgar:2022jxm}, where the entropy is complex.) The Taub-NUT has a null Killing vector $\xi=\partial_t$ on the horizon $r_+$ where $f(r_+)=0$. The temperature and entropy are
\be
T= \fft{1}{4\pi r_+}\,,\qquad S=\pi (r_+^2 + n^2)\,.
\ee
However, these are not enough to determine the first law, before the mass is determined.

   We now apply the Wald formalism, with formulae given in Appendix \ref{app:wald}, to the
Taub-NUT. We first present the ``raw data:''
\bea
&&{\mathbf{Q}} [\xi]= V(r) \Omega_\2 + U(r) dr\wedge (dt + 2n \cos\theta\, d\phi)\,,\qquad \Omega_\2=\sin\theta
d\theta\wedge d\phi\,,\nn\\
&&\delta {\mathbf Q}[\xi] - i_\xi {\mathbf \Theta} = \delta U(r)\, dr\wedge dt + X(r)\, \Omega_\2 +
Y(r)\, dr\wedge \cos\theta d\phi\,,\label{nutqtheta}
\eea
where
\bea
V &=& (r^2 + n^2) f'\,,\qquad U = \fft{2n f}{r^2 + n^2}\,,\nn\\
X &=& \fft{2}{r^2 + n^2}\Big(2n rf \delta n + (r^2 + n^2) (n f' \delta n- r\delta f)\Big)\,,\nn\\
Y &=& \fft{4n}{(r^2 + n^2)^2} \Big((3r^2 + n^2) f \delta n + n(r^2 + n^2) \delta f\Big)\,.
\eea
We see that the structures resemble those discussed earlier, involving either Dirac or Misner strings. For pure gravity, we have $d{\mathbf Q}[\xi]=0$, indicated by $V' + 2n U=0$. Following the same technique employed in the previous sections, we obtain the Smarr relation
\be
0=\fft{1}{8\pi} \int_{\Sigma} {\mathbf Q}[\xi] = \fft14 \Big(V + \int_{r_+}^r 2n U(r') dr'\Big)\Big|_{r_+}^\infty = \Big(m +\fft{n^2}{r_+}\Big) - 2T S\,.\label{nutsmarr0}
\ee
This suggests that we may define mass as
\be
M=m +\fft{n^2}{r_+}\,,\label{nutmass1}
\ee
and the first law then reads
\be
\delta M= T \delta S + \Phi_{\rm N} \delta Q_{\rm N}\,,\qquad \hbox{with}\qquad \Phi_{\rm N} = \fft{n}2\,,\qquad Q_{\rm N}=\fft{n}{r_+}\,.
\label{nutfl1}
\ee
In this interpretation, the NUT charge $Q_{\rm N}$ is a dimensionless parameter and therefore it does not enter the Smarr relation
\be
M=2TS\,.\label{nutsmarr1}
\ee
Note that after solving $m$ from $f(r_+)=0$, our mass \eqref{nutmass1} of the Taub-NUT becomes
\be
M=\fft12\big(r_+ + \fft{n^2}{r_+}\big)\ge |n|\,.
\ee
In other words, our mass is not only positive, but has a minimum $M_{\rm min}=n$, occurring when $m=0$. It is intriguing to observe that the Riemann-tensor squared vanishes on the horizon when the mass $M$ is minimum:
\be
R^{\mu\nu\rho\sigma}R_{\mu\nu\rho\sigma}\Big|_{r=r_+} = 0\,,\qquad\hbox{when $m=0$.}
\ee
However, the Riemann tensor is generally not zero on the horizon so it is not locally flat.

We now examine the Wald equality, originated from the integration of $d(\delta {\mathbf Q}[\xi] - i_\xi {\mathbf \Theta})=0$, guaranteed by $X' + Y=0$ in this particular example.  Integrating this out gives us
\bea
\fft{1}{16\pi} \int_\Sigma (\delta {\mathbf Q}[\xi] - i_\xi {\mathbf \Theta}) &=& \fft14 \Big(X(r) + \int_{r_+}^r Y(r') dr'\Big)\Big|_{r_+}^\infty\nn\\
&=& \Big(\delta m + \fft{3n}{2r_+} \delta n + \fft{n^2}{2} \delta (\fft{1}{r_+})\Big) -
T\delta S =0\,.\label{nutfoeq}
\eea
This first-order differential equation is certainly consistent with our statement of the first law \eqref{nutfl1}; however, \eqref{nutfoeq} does not give a unique or even an obvious choice of mass.  We may also define the mass as
\be
\widetilde M = m + \fft{n^2}{2r_+} = \fft{r_+}{2}\,,\label{nutmass2}
\ee
which is also nonnegative. Indeed this was recently adopted in \cite{Awad:2022jgn}. For this definition of energy we have
\be
\delta \widetilde M = T \delta S - Q_{\rm N} \delta \Phi_{\rm N}\,.
\ee
Thus we see that $\widetilde M = M - \Phi_{\rm N} Q_{\rm N}$ is the thermodynamic Legendre transformation from $M$. It is worth pointing out that both $M$ and $\widetilde M$ are positive definite for non-vanishing $n$, indicating that both are sensible candidates for the mass.

The free energy associated with the Euclidean action is of the Gibbs type if $M$ is the mass, and the Helmholts type if $\widetilde M$ is the mass:
\be
F=M - T S - \Phi_{\rm N} Q_{\rm N}  =\widetilde M - T S = \ft12m\,.
\ee
(Recall that the parameter $m$ can take all real values.) Thus we see that both definitions of the mass are consistent with the Euclidean action. Regardless the interpretation of mass,  the full differential of the free energy is uniquely determined:
\be
\delta F = -S \delta T - \fft{n}{2r_+} \delta n\,.
\ee
While this strongly suggests that $n$ is the thermodynamic potential, this is not the unique interpretation, since it is mathematically consistent to treat $m$ as the mass, and then one has $(\Phi_{\rm N}, Q_{\rm N})=(-1/(2n), n^3/r_+)$ \cite{Hennigar:2019ive}. The ambiguity in determining the first law originated from the fact that we do not have an independent check for any of the three quantities: mass, NUT charge and its thermodynamic potential. Thus it becomes a wild guess and this is not satisfactory to us, even if we make our preferred choice \eqref{nutmass1} without further rationale.

It is worth noting that $\mathbf Q[\xi]$ in \eqref{nutqtheta} also contains a time component. In other words, for a constant $\phi$ slice, we have
\be
\mathbf Q[\xi]=U(r) dr\wedge dt\,.
\ee
It is easy to see that
\be
\int i_\xi \mathbf Q[\xi] = \int_{r_+}^\infty U(r') dr' =\fft{n}{r_+}\,.\label{nutq}
\ee
This provides an explicit and concrete calculation for our NUT charge. The rationale is the following. In a vacuum solution where no source is provided, a charge such as the mass or the electric charge, is typically located at the singularity. Its quantity can be obtained from the Gaussian integration over a hypersurface that encloses the singularity. In other words, we need to integrate the full $u=\cos\theta \in [-1,1]$ coordinate region, and the outcome is then independent of the detail positions of the hypersurface, or the radial coordinate $r$. We may take a similar view of our NUT charge that it is also located at the singularity in the form of Misner strings, which stretch from the horizon to asymptotic infinity. It can be viewed that the Misner strings are precisely caused by the NUT charges distributed there. We thus integrate radially over, from $r_+$ to infinity, and the outcome is then independent of the latitudinal $u$. This parallel principle between $r$ and $u$ will guide us to obtain the NUT charges in later more complicated examples.

The result of the NUT charge however may appear counterintuitive since it is legitimate to question why the charge will depend on the existence of a horizon. As we pointed it out earlier, Taub-NUT has an usual property that for nonvanishing $n$, there is always the horizon. Thus the NUT charge not only creates the Misner strings, but also the event horizon! It is very important that unlike the Dirac strings discussed earlier, the Misner string is real and physical. The global change of the spacetime structure also implies that it can have consequence to the total mass. This is another difference from the Dirac strings that is not universal and neutral particles do not experience. For this reason, even though the math structures of the Dirac and Misner strings are the analogous in the Komar form $\mathbf Q[\xi]$, their contributions to the mass are treated differently. The above however does not explain why $n$ should be viewed as the thermodynamic potential, other than it is required by the satisfaction of the first law. Our definition of NUT charge becomes more apparent in the context of a larger solution with angular momentum, which we discuss next.

Before we progress to the next subsection, we would like to point out that our mass definition \eqref{nutmass1} can be also elegantly expressed as
\be
M=\sqrt{m^2 + n^2}\,,\label{nutmass3}
\ee
which indicates that $m$ and $n$ contributes equally to the total energy. This formula is analogous to the electric and magnetic charge contributions to the mass of the extremal RN black hole. It is suggestive that $(m,n)$ might be also viewed as ``gravitational electric and magnetic'' contributions to the mass. (This is distinct from viewing $n$ as the magnetic version of the mass.)

\subsection{Kerr-Taub-NUT}
\label{sec:kerr-taub-nut}

The Ricci-flat metric is given by
\bea
ds^2 &=& (r^2 + v^2) \Big(\fft{dr^2}{\Delta_r} + \fft{du^2}{1-u^2}\Big) +
\fft{1}{r^2 + v^2} \Big((1-u^2) e_1^2 + \Delta_r e_2^2 \Big)\,,\nn\\
e_1 &=& a dt - (r^2 + a^2 + n^2) d\phi\,,\qquad e_2 = dt + (2n u - a (1-u^2)) d\phi\,,\nn\\
\Delta_r &=& r^2 - 2mr + a^2 - n^2\,,\qquad v=n + a u\,,\qquad u=\cos\theta \in [-1,1]\,,
\label{ktnmetric}
\eea
The solution now contains three integration constants $(m,n,a)$. For $n=0$, the solution is the standard Kerr metric. We make a coordinate choice such that the metric is non-rotating at the asymptotic infinity, and it has an angular velocity $\Omega_+$ on the horizon where $\Delta_r(r_+)=0$. The corresponding null Killing vector is
\be
\xi = \partial_t + \Omega_+ \partial_\phi\,,\qquad \Omega_+ = \fft{a}{r^2 + a^2 + n^2}\,.
\label{ktnxi}
\ee
The temperature and entropy can be calculated straightforwardly, given by
\be
T=\fft{r_+^2 +n^2 - a^2}{4\pi r_+ (r_+^2 + n^2 + a^2)}\,,\qquad
S=\pi (r_+^2 + a^2 + n^2)\,.
\ee
To obtain the mass and angular momentum, we derive the 2-forms associated with the Komar integration (with fixed $t$):
\be
\mathbf Q[\partial_t]  = V(r,u) \Omega_\2 + U(r,u) dr\wedge d\phi\,,\qquad
\mathbf Q[\partial_\phi]  = X(r,u) \Omega_\2 + Y(r,u) dr\wedge d\phi\,,
\ee
where $\Omega_\2 = -du\wedge d\phi$ and
\bea
V&=&\fft{2(r^2 + a^2 + n^2)(2n r v + m (r^2 -v^2))}{(r^2 + v^2)^2}\,,\qquad
U=\fft{2(a^2 + n^2 - v^2)(2m r v - n (r^2-v^2))}{a(r^2 + v^2)^2}\,,\cr
X &=& \frac{2}{a \left(r^2+v^2\right)^2}\Big(m \left(a^2+n^2-v^2\right) \left(a^2 \left(v^2-r^2\right)+n^2 \left(v^2-r^2\right)-r^2 \left(3 r^2+v^2\right)\right)\cr
&&
-2 n r \left(2 n^2 v \left(a^2+r^2\right)+v \left(a^2+r^2\right)^2+n^4 v-n \left(r^2+v^2\right)^2\right)\Big)\,,\cr
Y &=& \frac{2}{a^2 \left(r^2+v^2\right)^2}
\Big(n (n-v)^2 \left(n^2 \left(r^2-v^2\right)+2 n v \left(r^2-v^2\right)+r^4+3 r^2 v^2\right)\cr
&&+n \left(a^4 \left(r^2-v^2\right)+a^2 \left(2 n^2 \left(r^2-v^2\right)+r^4+3 v^4\right)\right)-2 m r v \left(a^2+n^2-v^2\right)^2
\Big).\label{uvxy}
\eea
The closure of the Komar 2-forms are implied by the integrability conditions
\be
\partial_r V + \partial_u U =0\,,\qquad \partial_r X + \partial_u Y=0\,.\label{uvxycon}
\ee
The closure guarantees that integrating out the $u$ cycle gives a radial $r$ conserved quantity, and integrating out the $r$ cycle gives a latitudinal $\theta$ conserved quantity. Specially the radially conserved quantities can be constructed by
\bea
{\cal M}(r)&\equiv&\fft{\int d\phi}{8\pi}\Big( \int_{-1}^{+1} V(r,u) du + \int_{r_+}^r U(r',u)\Big|_{u=-1}^{u=+1} dr'\Big)= M=m + \fft{n^2}{r_+}\,,\nn\\
{\cal J}(r)&\equiv &-\fft{\int d\phi}{16\pi}\Big( \int_{-1}^{+1} X(r,u) du + \int_{r_+}^r Y(r',u)\Big|_{u=-1}^{u=+1} dr'\Big)=J=M a\,,
\eea
Note that $M$ is the same as in \eqref{nutmass1}. Note that for generic functions $(U,V)$ and $(X,Y)$, the quantities ${\cal M}(r)$ and ${\cal J}(r)$ would depend on $r$. They are in fact constants because of the integrability conditions \eqref{uvxycon}. In this definition of mass and angular momentum, the relation $J=Ma $ between the mass and angular momentum for the usual Kerr black hole is exactly retained, even when the NUT parameter is involved.  The mass can also be expressed as
\be
M=\frac{n^2 \sqrt{m^2+n^2-a^2}-a^2 m}{n^2-a^2}\,.
\ee
The elegant symmetry between $(m,n)$ in \eqref{nutmass3} is now broken by the rotation parameter $a$.

The first law is simply given by
\be
\delta M = T \delta S + \Omega_+ \delta J + \Phi_{\rm N} \delta Q_{\rm N}\,,\label{nutrotfl}
\ee
where the $(\Phi_{\rm N},Q_{\rm N})$ is the same as in \eqref{nutfl1} without rotation. The free energy from the Euclidean action is of the Gibbs type
\be
F_{\rm G}= M - T S - \Omega_+ J - \Phi_{\rm N} Q_{\rm N} = \ft12m\,.
\ee
We shall not present the details of the Wald formalism, since it will simply be consistent with the first law \eqref{nutrotfl}. In the extremal limit, the mass, angular momentum and the NUT charge are related by
\be
M_{\rm ext}^2 = |J| \sqrt{1 + Q_{\rm N}^2}\,.
\ee
In other words, the metric is black for $M\ge M_{\rm ext}$.

We now would like to have a deeper understanding of the NUT charge $Q_{\rm N}=n/r_+$, and explain why this is its suitable definition.  For both Taub-NUT and Kerr-Taub-NUT, in addition to the event horizon at $r=r_+$, there are two Killing horizons at north and south poles $(\theta=0,\pi)$, corresponding to $u=1$ and $u=-1$. The two degenerate Killing vectors are
\be
u=\pm 1:\qquad \ell_\pm = \partial_\phi \mp 2n \partial_t\,.\label{kerrnutell}
\ee
These two Killing vectors have unit ``Euclidean surface gravity'' \cite{Cvetic:2005zi}
\be
\kappa_{\rm E}^2=\fft{g^{\mu\nu}\partial_\mu \xi^2\partial_\nu \xi^2}{4\xi^2}=1\,.
\ee
Note that its definition does not have the minus sign that appears in the definition of the usual surface gravity \eqref{surfgrav}. In particular we have
\be
u=\pm 1:\qquad \kappa_{\rm E}=\pm 1\,.
\ee
The oppositive signs are analogous to the negative and positive temperatures associated with inner and outer horizons of a black hole. An event horizon is a hypersurface where the geodesic is not complete and matter can travel inside the horizon. The Killing horizon with Euclidean surface gravity on the other hand is geodesically complete and there is no connected spacetime behind a Killing horizon. Nevertheless the parallel of the two types of horizons, \eqref{kerrnutell} and \eqref{ktnxi} is striking, suggesting the correspondence \eqref{duality} mentioned in Introduction. Note that we view $\ell_\pm$ as given in \eqref{kerrnutell}, instead of $\partial_t \pm 1/(2n) \partial_\phi$ simply because \eqref{kerrnutell} has a smooth $n\rightarrow 0$ limit, and furthermore, when $n=0$, $\phi$ is the standard longitudinal coordinate, just as $t$ is the standard time in \eqref{ktnxi}.

Now the picture becomes clear. The angular momentum $J$, which is conjugate to $\Omega_+$, is obtained from the constant $t$ slice of the 2-form $\mathbf Q[\partial_\phi]$. By the duality correspondence, we expect that the $Q_{\rm N}$ is associated with the constant $\phi$ slice of the 2-form $\mathbf Q[\partial_t]$. For a constant $\phi$ slice, we have
\bea
&&{\mathbf Q}[\partial_t] = \big(\zeta(r,u) dr - \eta(r,u) du\big)\wedge dt\,,\label{Qt2form}\\
\zeta &=& \frac{2 \left(n \left(r^2-v^2\right)-2 m r v\right)}{\left(r^2+v^2\right)^2}\,,\qquad
\eta=-\frac{2 a \left(m \left(r^2-v^2\right)+2 n r v\right)}{\left(r^2+v^2\right)^2}\,.\nn
\eea
The closure of the 2-form ${\mathbf Q}[\partial_t]$ is ensured by the integrability condition
\be
\partial_u \zeta + \partial_r \eta=0\,.\label{zetaetaid}
\ee
By integrating out the $r$ coordinate, this allows us to obtain the $u$-invariant quantity, associated with Misner singularity at north and south poles $u=\pm 1$:
\bea
Q_{\rm N}^\pm = \fft12\Big( \int_{r_+}^\infty \zeta(r',u) dr' + \int_{\pm 1}^u \eta(r,u') \Big |_{r_+}^\infty du'\Big)= \fft{n\mp a}{2r_+}\,.\label{qnstep1}
\eea
Note that for generic functions $\zeta(r,u)$ and $\eta(r,u)$, the above $Q_{\rm N}^\pm$ quantities will depend on the variable $u$. However, owing to \eqref{zetaetaid}, they are constants.  The NUT charge is then given by
\be
Q_{\rm N} = Q_{\rm N}^+ + Q_{\rm N}^- = \fft{n}{r_+}\,.\label{qnstep2}
\ee
The reason for the sum above will be clear presently. The $dt$ factor in the 2-form \eqref{Qt2form} implies that $Q_{\rm N}$ is the growth rate of $u$-invariant time-like cycle of the 2-form $\mathbf Q[\partial_t]$.  Note that we have chosen the above normalization without a specific justification and consequently $\Phi_{\rm N}=n/2$. Turns off the angular momentum by setting $a=0$, the $Q_{\rm N}$ calculation becomes simply \eqref{nutq}.

It is worth remarking that while the surface gravity on the event horizon specifies the period of the Euclideanized time, the Euclidean surface gravity of the Killing horizon specifies the period of the longitudinal $\phi$. We have two Killing horizons with unit Euclidean surface gravity, and therefore both $\ell_\pm$ should generate $2\pi$ period. This implies that we must have $\Delta t = 4\pi n$ in order for the solution to be free from singularity. If we do so, it is effective making a coordinate change $t\rightarrow n t$, and $n$ becomes the periodicity of time. In our Taub-NUT approach, we must not impose this periodic time condition. We shall treat the time as a real line and consequently the solution has singularity at both north and south poles in the form of Misner strings. As we saw in the example of Kaluza-Klein monopole, making $t$ periodic changes the global structure completely and we shall not entertain the idea in this paper.

\subsection{Asymmetric Misner strings}
\label{sec:asym}

Up until now, we have been choosing the coordinates such that the Misner strings are attached to the north and south poles in the symmetric fashion. We can make a linear coordinate transformation among the Killing directions:
\be
t\rightarrow t - 2n \alpha \phi\,,\qquad \phi\rightarrow \phi\,,\label{asymt}
\ee
where $\alpha$ is a real numerical constant. Such a coordinate transformation is not allowed when $n=0$, since it will change the standard asymptotic Minkowski spacetime. For this reason, we multiply a factor $n$ in front of $\alpha$. Under this transformation, the degenerate Killing vectors at $u=\pm 1$ change to:
\be
u=\pm 1:\qquad  \ell_\pm = \partial_\phi \mp 4 \Phi_{\rm N}^\pm\,\partial_t\,,\qquad
\Phi_{\rm N}^\pm =\ft12 n (1\pm \alpha )\,.
\ee
Note that our sign choice in $\ell_\pm$ is related to the fact that in the vicinity of the degenerate cycles $u=1$ and $u=-1$, the coordinate $u$ can only either decrease or increase respectively. When the dimensionless parameter $\alpha=0$, it reduces to the special symmetric case discussed earlier. When $\alpha=1$, the Misner string disappears at the north pole, and it is likewise at the south pole when $\alpha=-1$. This implies that there is really only one independent Misner string. We shall see presently that this has a consequence on $\alpha$ in its role in the thermodynamic first law.

As we have mentioned earlier, when $n$ is nonvanishing, the spacetime is not asymptotic to the Minkowskian, and therefore there is no apparent fiducial $\alpha$ such that the Misner strings can be both removed. Consequently, the mass, angular momentum {\rm etc.}, have nontrivial dependence on the parameter $\alpha$. Thus the general $\alpha$ case provides a strong test for our procedure, since by this stage, all our thermodynamic quantities follow a strict and hands off the steering wheel procedure to calculate. We shall not present the detail calculation, since it will be repetitive. We present only the results:
\bea
M&=& m + 2\Phi_{\rm N}^+ Q_{\rm N}^+ + 2\Phi_{\rm N}^- Q_{\rm N}^-\,,\qquad J=M \Big( a + (\Phi_{\rm N}^--\Phi_{\rm N}^+)\Big)\,,\nn\\
T&=& \fft{r_+^2 + n^2-a^2}{4\pi r_+ (r_+^2 + n^2 + a^2 + 2\alpha n a)}\,,\qquad
S=\pi (r_+^2 + n^2 + a^2 + 2\alpha n a)\,,\nn\\
\Omega_+ &=& \fft{a}{r_+^2 + n^2 + a^2 + 2\alpha n a}\,,\qquad \Phi_{\rm N}^\pm =\ft12 n (1\pm \alpha )\,,\qquad Q_{\rm N}^\pm=\fft{n\mp a}{2r_+}\,.
\eea
Note that all except $Q_{\rm N}^\pm$ depend on the parameter $\alpha$ nontrivially. It is now straightforward to verify that the first law holds, namely
\be
\delta M = T \delta S + \Omega_+ \delta J + \Phi_{\rm N}^+ \delta Q_{\rm N}^+
+ \Phi_{\rm N}^- \delta Q_{\rm N}^-\,.
\ee
What is perhaps surprising is that the above first law is valid even if we treat the dimensionless parameter $\alpha$ as a thermodynamical variable, i.e.~$\delta \alpha\ne 0$. It plays a role of splitting the degeneracy of the NUT charges associated with two Misner strings at the north and south poles. Thermodynamically it is consistent to set $\alpha=0$ to a reduced system, and we recover the previous result with the intensive potential $\Phi_{\rm N}=\Phi^\pm_{\rm N}$ and extensive charges $Q_{\rm N}=Q_{\rm N}^+ + Q_{\rm N}^-$. This explains the sum rule of \eqref{qnstep2}.

It is interesting to observe that except for $J$ and $\Phi_{\rm N}^\pm$, there is no thermodynamic asymmetry of the two Misner strings under the coordinate transformation \eqref{asymt} when the angular velocity vanishes ($a=0$.) The asymmetry is actually promoted by the angular velocity and interestingly the angular momentum can be non-zero even when we turn off the angular velocity. Specifically, when $a=0$, although we have $\Omega_+=0$, we have an intriguing relation
\be
J=M (\Phi_{\rm N}^--\Phi_{\rm N}^+)\,.
\ee

Finally, as we should expect, the Smarr relation $M=2TS + 2 \Omega_+ J$ remains true for general $\alpha$. The final important test is that Euclidean action should not depend on the coordinate transformation, and the free energy should be independent on $\alpha$. Indeed we have
\be
F_G=M - T S -\Omega_+ J- \Phi_{\rm N}^+ Q_{\rm N}^+-\Phi_{\rm N}^- Q_{\rm N}^-=\ft12m\,.
\ee
This provides a strong evidence of the validity of our approach.

We therefore have obtained all the four ``charges:'' the mass $M$, angular momentum $J$ and the NUT charges $Q_{\rm N}^\pm$, associated with the four independent parameters $(m,a,n,\alpha)$ of the Kerr-Taub-NUT metric. We would like now to continue the comments at the end of the last subsection. We followed the principle outlined there to get these charges: The mass and angular momentum are the radially independent quantities after integrating out the $u$-cycles of ${\mathbf Q}[\partial_t]$ and ${\mathbf Q}[\partial_\phi]$. The NUT charges on the other hand spread along the Misner strings in the radial direction and they are the $u$-independent quantities after integrating over the $r$ cycle of ${\mathbf Q}[\partial_t]$. Here we would like to summarize the technicality of how to evaluate these quantities. For simplicity, we focus on the relevant 1-form
\be
\Xi=X(r,u)\, du - Y(r,u)\, dr\,.
\ee
We assume that it is close, $d\Xi=0$, which implies that there exists a scalar quantity $\Upsilon(r,u)$ such that $\Xi=d\Upsilon$. Therefore, we have $X=\partial_u \Upsilon$ and $Y=-\partial_r \Upsilon$. (The minus was added to be consistent with our earlier convention.) There exist two $u$-independent quantities
\be
-\Big( \int_{r_+}^\infty dr'\, Y(r',u)  + \int_{\pm 1}^u du'\, X(r,u') \Big |_{r_+}^\infty \Big)=\Upsilon(u=\pm 1)\Big|_{\infty}^{r_+}\,,\label{techniqueu}
\ee
and one $r$-independent quantity
\be
\int_{-1}^{+1} du X(r,u)+ \int_{r_+}^r dr'\, Y(r',u)\Big|_{u=-1}^{u=+1} =
\Upsilon(r_+)\Big|_{u=-1}^{u=1}\,.\label{techniquer}
\ee

\section{The Plebanski solution}
\label{sec:pleb}

The nontrivial test of our interpretation of the NUT charge is to apply the procedure to the general Plebanski metrics to obtain first all the thermodynamic quantities and then verify the first law.  The general Plebanski solution is very complicated, containing both the Dirac and Misner strings. It is a solution in Einstein-Maxwell gravity
\be
{\cal L} = \sqrt{-g} (R - F^2)\,,\qquad F_\2=dA_\1\,.\label{emlag}
\ee
(We shall not consider the cosmological constant in this paper.) Note that out of respect, we adopt the same convention for the kinetic term of the Maxwell field as in \cite{Plebanski:1975xfb}, without the $1/4$ factor that we used for the EMD theories.

\subsection{The solution and global analysis}

The Plebanski solution is of cohomogeneity two, depending on both the radial and latitude angular coordinates, as in the case of the Kerr black hole. The original solution was written in an elegant form with symmetric radial and latitudinal coordinates $(q,p)$. The general type-D solution to \eqref{emlag} is
\bea
ds^2 &=& (p^2 + q^2)\Big(\fft{dp^2}{P(p)} + \fft{dq^2}{Q(q)}\Big) +
\fft{1}{p^2 + q^2} \Big(P(p) \sigma_q^2 - Q(q) \sigma_p^2\Big)\,,\nn\\
A_\1 &=& \fft{1}{p^2 + q^2} (e q \sigma_p + g p \sigma_q)\,,\qquad B_\1=\fft{1}{p^2 + q^2}
(g q \sigma_p - e p \sigma_q)\,.
\eea
where
\be
P(p)= b - g^2 + 2 n p - \epsilon p^2\,,\qquad
Q(q)= b + e^2 - 2m q + \epsilon q^2\,,
\ee
and $\sigma_p$ and $\sigma_q$ are the 1-forms
\be
\sigma_p = d\tau - p^2 d\sigma\,,\qquad \sigma_q=d\tau + q^2 d\sigma\,.
\ee
Note that we presented above not only the Maxwell field $A_\1$, but also the gauge potential of the Hodge dual field strength $\widetilde F_\2=dB_\1 = {*F}_\2$, with the convention
\be
{*F}_{\mu\nu}=\ft{1}{2}\epsilon_{\mu\nu\rho\sigma}F^{\rho\sigma}.
\ee
The electromagnetic duality is given by
\be
F_\2\rightarrow {*F}_\2\,,\qquad {*F}_\2\rightarrow-F_\2\,,\label{emdual}
\ee
implemented in the Plebanski solution by $e\rightarrow g$ and $g\rightarrow -e$. The minus sign is consistent with the fact that there can be no self-duality for the Maxwell field strength in four dimensions with the Lorentzian signature, i.e.~$F_\2 \ne \pm {*F}_\2$.

The Plebanski solution appears to have six integration constants $(m,n,e,g,b,\epsilon)$ and the metric is flat when four of them $(m,n,e,g)=0$. The solution is invariant under the constant scaling
\bea
&&(p,q)\rightarrow \lambda (p,q)\,,\qquad \tau\rightarrow \fft{\tau}{\lambda}\,,\qquad
\sigma\rightarrow \fft{\sigma}{\lambda^3}\,,\nn\\
&&(e,g,\epsilon)\rightarrow \lambda^2 (e,g,\epsilon)\,,\qquad (m,n)\rightarrow \lambda^3 (m,n)\,,\qquad b\rightarrow \lambda^4 b\,.
\eea
We therefore can set without loss of generality $\epsilon=\pm 1,0$. (In the most general type-D Plebanski-Demianski solution \cite{Plebanski:1976gy}, an overall conformal factor breaks the scaling symmetry and the parameter $\epsilon$ becomes nontrivial there describing the acceleration of the black holes.) In this paper, we consider $\epsilon=1$. With this choice, it is clear that $q$ is noncompact and we rename it as the more familiar radial coordinate $r$, i.e.~$q=r$. On the other hand, coordinate $p$ is compact, and we introduce the familiar latitudinal coordinate $\theta$ by
\be
p=n + a u\equiv v\,,\qquad u=\cos\theta\,,\qquad a=\sqrt{b-g^2+n^2}\,.
\ee
Note that these are the same $(u,v)$ introduced in subsection \ref{sec:kerr-taub-nut}.  The radial profile function $Q(q)$ now becomes
\be
Q(q)\rightarrow \Delta_r = r^2 - 2m r + a^2+e^2+g^2-n^2\,.\label{plebdeltar}
\ee
We further make a linear coordinate transformation in the two Killing directions, namely the time and longitudinal angle:
\be
\tau=t - \fft{a^2 + n^2}{a} \phi\,,\qquad \sigma = -\fft{1}{a} \phi\,.
\ee
We now arrive at the metric that takes the identical form as that of Kerr-Taub-NUT \eqref{ktnmetric}, but with $\Delta_r$ given by \eqref{plebdeltar}. The Maxwell field $A_\1$ and the gauge potential $B_\1$ of the dual field strength are
\be
A_\1=\frac{g (v-n)}{a \left(r^2+v^2\right)} e_1 +\frac{(e r+g n)}{r^2+v^2}e_2\,,\qquad
B_\1=\frac{e(n-v)}{a \left(r^2+v^2\right)} e_1 +\frac{ (g r-e n)}{r^2+v^2}e_2\,,
\ee
where the 1-forms $(e_1,e_2)$ are given by \eqref{plebdeltar}. We therefore arrive at the full solution in terms of the familiar four-dimensional $(t,r, \theta,\phi)$ coordinates.

The solution is asymptotic to Minkowski spacetime in a non-rotating frame at large $r$ if the NUT parameter $n$ vanishes; otherwise, the solution is asymptotic to the Kerr-Taub-NUT. There are three degenerated cycles, associated with the horizon $r=r_+$ where $\Delta_r(r_+)=0$, and also north and south poles $u=\pm1$.  The null Killing vector on the horizon is
\be
\xi = \partial_t + \Omega_+ \partial_\phi\,,\qquad \Omega_+ = \fft{a}{r_+^2 + a^2 + n^2}\,,
\ee
from which we obtain the angular velocity $\Omega_+$.  The temperature and entropy can be calculated in the standard way described earlier, and they are
\be
T=\fft{r_+^2+n^2 - a^2 - e^2 - g^2}{4\pi r_+(r_+^2 + a^2 + n^2)}\,,\qquad S=\pi (r_+^2 + a^2 + n^2)\,.
\ee
The degenerate Killing vectors on the two Killing horizons are again \eqref{kerrnutell}.
Both of them give the unit Euclidean surface gravity.  This implies that if we treat time as a real line, then there are singularities in the form of two Misner strings, one at the north pole and the other at the south pole. Since the $\ell_\pm$ takes the same form as the ones in Taub-NUT or Kerr-Taub-NUT metrics, we have
\be
\Phi_{\rm N}=\ft12 n\,.\label{emphin}
\ee

In order to obtain the first law for the Plebanski solution, it is necessary to obtain all the thermodynamical quantities.  The electric and magnetic potentials are relatively easy, given by
\be
\Phi_e= \xi^\mu A_\mu(r_+) = \fft{e r_+ + n g}{r_+^2 + a^2 + n^2}\,,\qquad
\Phi_m = \xi^\mu B_\mu(r_+) = \fft{g r_+- n e}{r_+^2 + a^2 + n^2}\,.\label{empot}
\ee
However, there is actually a subtlety here also.  The above calculation is in fact rather sloppy, since the quantity $\xi^\mu A_\mu$ is not gauge invariant. This usually can be solved by considering the potential difference, namely $\xi^\mu (A_\mu(r_+) - A_\mu (\infty))$, which however turns out to be dependent on the coordinate $u=\cos\theta$ for the general Plebanski solution
\be
\xi^\mu (A_\mu(r_+) - A_\mu (\infty))=\frac{e r_++n g}{r_+^2+ a^2+n^2}+\frac{a g u}{r_+^2+a^2+n^2}\,.
\ee
It should be emphasized the extra term does not arise because of the NUT parameter, but because of rotation and magnetic charge. This term actually exists even in the more straightforward dyonic Kerr-Newman black hole. In order to obtain the correct potentials that are independent of the gauge choice, we need to apply the formula
\bea
\Phi_e &=& \ft12 \xi^\mu \Big(A_\mu (u=1)+ A_\mu(u=-1)\Big)\Big|^{r_+}_{\infty}\,,\nn\\
\Phi_g &=& \ft12 \xi^\mu \Big(B_\mu (u=1)+ B_\mu (u=-1)\Big)\Big|^{r_+}_{\infty}\,.\label{potformula}
\eea
These formulae turn out to be rather universal in order to obtain the relevant electric and magnetic potentials associated with other null Killing vectors in the manifold, {\it e.g.}~$\ell_\pm$. The plus sign in the formulae are counterintuitive since one expects the difference. The plus sign is originated from the fact that the Euclidean surface gravities at $u=\pm 1$ Killing horizons have oppositive signs.

\subsection{Thermodynamics}

It is a tedious but straightforward process to verify the Wald 2-form is indeed close, i.e.,
\be
d(\delta {\mathbf Q}[\xi] - i_\xi {\mathbf \Theta}) =0\,.
\ee
However, with the solution involving five parameters $(m,n,a,e,g)$, it is not simple to extract the relevant thermodynamic quantities to give a precise statement of the first law. Following the Kerr-Taub-NUT example, we can first read off the mass, angular momentum and NUT charges from the closed 2-form $\mathbf Q$.  However when matter is involved, the $\mathbf Q$ defined by Wald is not closed. When Misner strings are involved, we can no longer simply read off these charges by integrating $\mathbf Q$'s over the sphere at asymptotic infinity. We therefore must have the closed 2-form $\mathbf Q$. The closed generalized Komar 2-form is given by
\be
\widetilde {\mathbf Q}[\xi] = \fft1{4} \epsilon_{\alpha\beta \mu\nu} \widetilde Q^{\alpha\beta} dx^\mu \wedge dx^\nu\,,
\ee
where for Einstein-Maxwell gravity \eqref{emlag}, we have
\be
\widetilde Q^{\mu\nu}=-2\nabla^{[\mu}\xi^{\nu]}-4 F^{\mu\nu}A_{\lambda}\xi^\lambda + 4
F^{[\mu |\rho|} \xi^{\nu]} A_\rho\,.
\ee
This quantity does not have the electromagnetic duality. We can add an additional closed 2-form $\fft12 d((i_\xi B_\1) A_\1)$ to $\widetilde {\mathbf Q}$ so that the resulting one has the duality.  We find that
\be
\widetilde Q^{\mu\nu}=-2\nabla^{[\mu}\xi^{\nu]}-2 F^{\mu\nu}A_{\lambda}\xi^\lambda - 2 \widetilde F^{\mu\nu}B_{\lambda}\xi^\lambda
\ee
is both closed and invariant under the electromagnetic duality.

We are now in the position to derive the mass, angular momentum and the NUT charge.  It turns out that the $\widetilde {\mathbf Q}$ takes the  same form as the one given in the Kerr-Taub-NUT cases. For example, we have
\bea
\widetilde {\mathbf Q}[\partial_t] &=& {\mathbf Q}[\partial_t]_{\rm Kerr-Taub-NUT}\nn\\
 &=& V(r,u) \Omega_\2 + U(r,u)\,dr\wedge d\phi + \big(\zeta(r,u) dr - \eta(r,u) du\big)\wedge dt\,,
\eea
where the functions of all components are exactly the same as those given in \eqref{uvxy}. The same story also goes with $\widetilde {\mathbf Q}[\partial_\phi]$. These results are understandable, since metrics of the Kerr-Taub-NUT and Plebanski take the same form except that the function $\Delta_r$ is different. In the Plebanski solution, the Wald 2-form ${\mathbf Q}$ is therefore no longer close. The extra terms in $\widetilde {\mathbf Q}$ has precisely the effect to removing the extra $(e,g)$ contributions that enter $\Delta_r$ in the Plebanski solution such that the $\widetilde {\mathbf Q}$ now has the same expression as ${\mathbf Q}_{\rm Kerr-Taub-NUT}$ which is closed.  Following the same technique, we have
\bea
M= m + n Q_{\rm N}\,,\qquad J= M a\,,\qquad
Q_{\rm N}= \fft{n}{r_+}\Big(1 - \fft{(e^2 + g^2)(r_+^2 + n^2-a^2)}{(r_+^2 + n^2 + a^2)^2 - 4 n^2 a^2}\Big)\,.
\label{plebqn}
\eea
These results are numerically differently from those of Kerr-Taub-NUT because the location of the horizon is modified by the electric and magnetic charge parameters $(e,g)$.

The electric and magnetic charges of the solutions with the NUT parameter involved can be also subtle to evaluate, compared to the dyonic Kerr-Newman solution. From the Maxwell equation $d\widetilde F_\2=0$ and the Bianchi identity $dF_\2=0$, we see that the electric and magnetic charges are related to the integrations of the closed 2-forms $\int \widetilde F_\2$ and $\int F_\2$ respectively. The technique is then same as we extracted the charges from the closed $\mathbf Q$ or $\widetilde {\mathbf Q}$ forms. Specifically, the electric charge is related to the $r$-independent cycle
\bea
Q_e &=& -\fft12\Big(\int_{-1}^1 \widetilde F_{u\phi}(r,u') du' +
\int_{r_+}^r \widetilde F_{r\phi}(r',u)\Big|_{u=-1}^{u=1} dr'\Big)\nn\\
&=&\ft12 B_\phi(r_+) \Big|_{u=-1}^{u=1} = e + 2 n Q_{e{\rm N}}\,.
\eea
Likewise, the magnetic charge is now given by
\be
Q_g= - \ft12 A_\phi(r_+)\Big|_{u=-1}^{u=1} = g - 2 n Q_{g{\rm N}}\,,
\ee
where
\bea
Q_{e{\rm N}} &=& \fft{g r_+ \left(r_+^2+n^2+a^2\right)-e n \left(r_+^2 + n^2-a^2\right)} {(r_+ + n^2+a^2)^2 - 4 n^2 a^2}\,,\nn\\
Q_{g{\rm N}} &=& \fft{e r_+ \left(r_+^2 + n^2 + a^2\right)+g n \left(r_+^2 + n^2-a^2\right)}
{(r_++n^2+a^2)^2 - 4 n^2 a^2}\,.
\eea
In the above calculations, we have considered the constant time slice of the $F_\2$ and $\widetilde F_\2$. The integration of $\phi$ is straightforward, giving to $2\pi$. The remaining one-dimensional integration follows the technique described in \eqref{techniquer}.

With these thermodynamic variables, we find that the expected Smarr relation is satisfied, namely
\be
M = 2T S + 2 \Omega_+ J + \Phi_e Q_e + \Phi_g Q_g\,.\label{plebsmarr}
\ee
But this is not yet the end of story.  Recall that in addition to the null Killing vector $\xi$, we also have the two degenerate Killing vectors $\ell_\pm$ associated with the Misner strings at the north and south poles. Each Misner string can induce electric and magnetic NUT charges and their conjugate potentials.  This leads to a total of four electromagnetic NUT charges.  Following the same technique we developed in \eqref{techniqueu}, we find the four charges are
\bea
Q_{e{\rm N}}^{\pm} &=& \ft12 B_t (u=\pm1)\Big|_\infty^{r_+}=
\fft{gr_+ -e n \mp a e}{2(r_+^2 + (n\pm a)^2)}\,,\nn\\
Q_{g{\rm N}}^{\pm} &=& \ft12 A_t (u=\pm1)\Big|_\infty^{r_+}= \fft{er_+ + g n \pm a g}{2(r_+^2 + (n\pm a)^2)}\,.
\eea
Their conjugate potentials can be obtained following the analogous rules \eqref{potformula} for the electric and magnetic potentials. We find
\bea
\Phi_{e{\rm N}}^{\pm} &=& \pm \ft18 \ell^\mu_\pm
\Big(A_\mu(u=1) + A_\mu(u=-1)\Big)\Big|^{r_+}_{\infty}= -\fft{n(er_+ + gn \mp a g)}{r_+^2 + (n\mp a)^2}\,,\nn\\
\Phi_{g{\rm N}}^{\pm} &=& \mp \ft18 \ell^\mu_\pm
\Big( B_\mu(u=1) +  B_\mu(u=-1)\Big)\Big|^{r_+}_{\infty}= \fft{n(g r_+ - e n \pm ae)}{r_+^2 + (n\mp a)^2}\,.
\eea
It is worth pointing out some intriguing relations
\be
Q_e = e + 4\Phi_{\rm N} (Q_{e{\rm N}}^{+} + Q_{e{\rm N}}^{-})\,,\qquad
Q_g = g - 4\Phi_{\rm N} (Q_{g{\rm N}}^{+} + Q_{g{\rm N}}^{-})\,.
\ee
In other words, we have $Q_{e{\rm N}}=Q_{e{\rm N}}^{+} + Q_{e{\rm N}}^{-}$ and
$Q_{g{\rm N}}=Q_{g{\rm N}}^{+} + Q_{g{\rm N}}^{-}$. We find that the first law takes the form
\bea
\delta M &=& T \delta S + \Omega_+ \delta J + \Phi_e \delta Q_e + \Phi_g \delta Q_g\nn\\
&& +
\Phi_{\rm N} \delta Q_{\rm N} + \Phi_{e{\rm N}}^{+} \delta Q_{e{\rm N}}^{+} + \Phi_{e{\rm N}}^{-} \delta Q_{e{\rm N}}^{-} + \Phi_{g{\rm N}}^{+} \delta Q_{g{\rm N}}^{+}+ \Phi_{g{\rm N}}^{-} \delta Q_{g{\rm N}}^{-}\,.
\eea
The terms in the second line are all associated with the NUT-related charges.  Since these charges are all dimensionless, they do not affect the Smarr relation \eqref{plebsmarr}.

The final test is whether these are consistent with the free energy from the Euclidean action. We find
\bea
F &=& M - T S - \Omega_+ J - \Phi_e Q_e - \Phi_{\rm N} Q_{\rm N} - \Phi_{e{\rm N}}^{+}  Q_{e{\rm N}}^{+} - \Phi_{e{\rm N}}^{-} Q_{e{\rm N}}^{-}\cr
&=& \ft12 m - \fft{r_+ ((e^2-g^2) \left(r_+^2+ a^2-n^2\right)+4 e g n r_+)}{2((r_++n^2+a^2)^2 - 4 n^2 a^2)}\,.\label{plebnutfree}
\eea
This is indeed the one from the Euclidean action. Note that the Legendre transformation is associated with all the electrically related charges, with the magnetic charges uninvolved, analogous to the Kerr-Newmann black hole. Note also that although $Q_{e\rm N}^\pm$ and $Q_{g\rm N}^\pm$ are nonvanishing when $n=0$, their conjugate potentials do vanish, such that they disappear from the first law or the Euclidean action in the absence of the NUT charge.

Finally the first law has the electromagnetic duality, simply realized by
\be
e\rightarrow g\,,\qquad g\rightarrow -e\,.\label{egtrans}
\ee
This expression of the duality is a consequence of the convention \eqref{emdual}. Such manifest electromagnetic duality was absent in the previous approach to the thermodynamics for the general Plebanski solution.

\subsection{Asymmetric Misner strings}

We can now go the Full Monty and consider the Plebanski solution with asymmetric Misner strings, arising from the coordinate transformation \eqref{asymt}. We shall not repeat our well-defined procedures, but simply present all the thermodynamic variables
\bea
&&M=m + 2 \Phi_{\rm N}^+ Q_{\rm N}^+ + 2\Phi_{\rm N}^- Q_{\rm N}^-\,,\nn\\
&&J=M \Big(a + (\Phi_{\rm N}^- - \Phi_{\rm N}^+)\Big) + \fft{\alpha n r_+(e^2+g^2)(r_+^2+ n^2-a^2)}{(r_+^2 + n^2+a^2)^2 - 4 n^2 a^2}\,,\nn\\
&&T = \fft{r_+^2+ n^2 -a^2 - e^2 -g^2}{4\pi r_+(r_+^2 + n^2 + a^2 + 2\alpha a n)}\,,\qquad
S=\pi (r_+^2 + n^2 + a^2 + 2\alpha a n)\,,\nn\\
&&\Omega_+ = \fft{a}{r_+^2 + n^2 + a^2 + 2\alpha a n}\,,\qquad \Phi_{\rm N}^\pm=\ft12 n(1\mp \alpha)\,,\nn\\
&&Q_{\rm N}^\pm = \fft{n\mp a}{2r_+} - \fft{(e^2+g^2)\big(n(n^2+r_+^2-a^2)\pm a (3r_+^2+a^2-n^2)\big)} {2r_+ ((r_+^2 + n^2+a^2)^2 - 4 n^2 a^2)}\,,\nn\\
&&\Phi_e = \fft{e r_+ + g n}{r_+^2 + n^2 + a^2 + 2\alpha a n}\,,\qquad
\Phi_g = \fft{g r_+ - e n}{r_+^2 + n^2 + a^2 + 2\alpha a n}\,,\nn\\
&&Q_e = e + 4\Phi_{\rm N}^+ Q_{e{\rm N}}^+ +4\Phi_{\rm N}^- Q_{e{\rm N}}^-\,,\qquad
 Q_g = g - 4\Phi_{\rm N}^+ Q_{g{\rm N}}^+ - 4\Phi_{\rm N}^- Q_{g{\rm N}}^-\,,\nn\\
&&Q_{e{\rm N}}^\pm = \fft{gr_+ - e (n\pm a)}{2(r_+^2 + (a\pm n)^2)}\,,\qquad
Q_{g{\rm N}}^\pm = \fft{er_+ + g (n \pm a)}{2(r_+^2 + (a \pm n)^2)}\,,\nn\\
&&\Phi_{e{\rm N}}^\pm = -\fft{n(er_+ + g(n\mp a))}{r_+^2 + (n\mp a)^2}\,,\qquad
\Phi_{g{\rm N}}^\pm = \fft{n(g r_+ - e (n\mp a))}{r_+^2 + (n\mp a)^2}\,.
\eea
Note that the NUT induced electromagnetic charges and their potentials, i.e.~the variables in the last two lines, are all independent of $\alpha$. It is interesting to note that when $\alpha n\ne 0$, the angular momentum is nonvanishing even with zero angular velocity $(a=0)$. It is straightforward to verify that the complicated first law holds, namely
\bea
\delta M &=& T \delta S + \Omega_+ \delta J + \Phi_e \delta Q_e + \Phi_g \delta Q_g\nn\\
&& + \Phi_{\rm N}^+ \delta Q_{\rm N}^+ +\Phi_{\rm N}^- \delta Q_{\rm N}^- + \Phi_{e{\rm N}}^{+} \delta Q_{e{\rm N}}^{+} + \Phi_{e{\rm N}}^{-} \delta Q_{e{\rm N}}^{-} + \Phi_{g{\rm N}}^{+} \delta Q_{g{\rm N}}^{+}+ \Phi_{g{\rm N}}^{-} \delta Q_{g{\rm N}}^{-}\,.
\eea
As in the Kerr-Taub-NUT case with asymmetric Misner strings, although many thermodynamic variables have nontrivial dependence on $\alpha$, both $\alpha$ and $\delta \alpha$ drop out miraculously from the above first law even if we treat it as a variable rather than a constant, indicating that $\alpha$ is not a spurious parameter thermodynamically. The Smarr relation is still given by \eqref{plebsmarr}, unaffected by $\alpha$. The free energy from the Euclidean action should not depend on the trivial coordinate transformation \eqref{asymt}, and indeed we have
\be
F = M - T S - \Omega_+ J - \Phi_e Q_e - \Phi_{\rm N}^+ Q_{\rm N}^+ -\Phi_{\rm N}^- Q_{\rm N}^- - \Phi_{e{\rm N}}^{+}  Q_{e{\rm N}}^{+} - \Phi_{e{\rm N}}^{-} Q_{e{\rm N}}^{-}\,,
\ee
which is independent of $\alpha$, given by the second equality of \eqref{plebnutfree}. Our results indicate that knowing both the Smarr relation and the Euclidean action is still far from determining the thermodynamic first law. When $\alpha=0$, the $(\Phi_{\rm N}, Q_{\rm N})$ of \eqref{emphin} and \eqref{plebqn} are given by $\Phi_{\rm N}=\Phi_{\rm N}^\pm$ and $Q_{\rm N} =Q_{\rm N}^+ + Q_{\rm N}^-$, respectively. Our results however raise a puzzling aspect that we have more thermodynamic variables than the total number of the parameters in the solution. This seems to suggest that there may exist a more general class of solutions where these NUT induced electromagnetic charges are independent parameters. Analogous situation occurs in the Kuluza-Klein dyonic AdS black hole in gauged supergravity \cite{Lu:2013ura} where the completion of the first law requires an extra $XdY$ that is associated with the scalar hair contribution even though there is no such an independent parameter in the exact solution; the independent scalar hair parameter emerges in the more general class of numerical solutions.
It is worth remarking here that two candidates for the first law with five independent thermodynamic variables were obtained, by introducing electric or magnetic charges on the horizon \cite{BallonBordo:2020mcs}. The resulting first law was nicely compact, but it suffers from two inadequacies: lacking of a smooth $n\rightarrow 0$ limit and the absence of electromagnetic duality despite the fact that the local solution has it.

Finally we would like to present the explicit transformations of all the thermodynamic variables under the discrete electromagnetic duality transformation \eqref{egtrans}. All the neutral variables $(M, J,  \Omega_+, T, S, Q_{\rm N}^\pm, \Phi_{\rm N}^\pm)$ are invariant and the charged variables transform as
\bea
&&\Phi_e\rightarrow \Phi_g\,,\qquad \Phi_g\rightarrow - \Phi_e\,,\qquad
Q_e\rightarrow Q_g\,,\qquad Q_g\rightarrow -Q_g\,,\nn\\
&&Q_{e{\rm N}}^\pm \rightarrow -Q_{g{\rm N}}^\pm \,,\qquad Q_{g{\rm N}}^\pm \rightarrow
Q_{e{\rm N}}^\pm\,,\qquad
\Phi_{e{\rm N}}^\pm \rightarrow -\Phi_{g{\rm N}}^\pm \,,\qquad \Phi_{g{\rm N}}^\pm \rightarrow
\Phi_{e{\rm N}}^\pm\,.
\eea
There can be no self-duality.

\section{Conclusions}
\label{sec:con}

Our goal is to provide well defined procedures of calculating all the thermodynamic variables in the Taub-NUT spacetime and its Plenbanski generalization, and then verify the first law of their black thermodynamics. We assume that the standard procedures to obtain the Hawking temperature and Bekenstein-Hawking entropy are valid. However, the mass, NUT charge and its thermodynamic conjugate potential are not easy to determine. This is because the Taub-NUT metric is not asymptotic to the Minkowski spacetime and the usual ADM mass does not apply. There is neither an obvious definition of the so called ``NUT-charge.'' In order to decipher these quantities, we used a variety of tools including the Euclidean action approach, Wald formalism and Komar and generalized Komar integrations.

The application of these tools in the Taub-NUT geometries can be subtle. In particular, the Plebanski solution involves both Dirac and Misner singularities. We therefore set to develop the formalism using both electric and magnetic black holes in a class of EMD theories.  We also studied the same issues on the five-dimensional boosted string and Kaluza-Klein monopole. These exercises enabled us finally to decipher the spacetime structures of Taub-NUT and Plebanski solutions and obtain all of their thermodynamic variables and verify the first law.

The key in our approach is to treat the time as a real line so that the Misner string singularties are real and physical. We found that the NUT charge spreads along the Misner strings, just as the mass of the Schwarzschild black hole is located at its spacetime singularity. Our technic breakthrough was the observation of the parallel between the null Killing vector $\xi$ on the horizon \eqref{nullxi0} and the degenerate Killing vectors $\ell_\pm$ at the north and south poles \eqref{ellpm0}.  This suggests the correspondence \eqref{duality}, which not only identifies the NUT potential as being proportional to the parameter $n$, but also provides us a way of calculating the NUT charge from the closed Komar or generalized Komar form $\widetilde {\mathbf Q}$. Just as we can obtain the angular momentum as the radially-conserved quantity from the constant-$t$ slice of $\widetilde {\mathbf Q}[\partial_\phi]$ by integrating out the angular coordinates, we can obtain the NUT charge, up to a purely numerical scaling factor, as the angular-invariant quantity by integrating the constant-$\phi$ slice of $\widetilde {\mathbf Q}[\partial_t]$, radially from the horizon to asymptotic infinity. This leads to some salient features in our approach: The NUT charge and its potentials both have smooth $n\rightarrow 0$ limit. The mass is nonnegative for all black holes and the resulting thermodynamic charges are simpler compared to, e.g.~ \cite{BallonBordo:2019vrn}, for the Kerr-Taub-NUT black hole.

The parallel continues when Maxwell fields are included.  Just as the electric and magnetic potentials are associated with $\xi^\mu A_\mu$ and $\xi^\mu B_\mu$, we expected that there were four NUT-induced electric and magnetic potentials associated with $\ell_\pm^\mu A_\mu$ and $\ell_\pm^\mu B_\mu$. The electric and magnetic charges are the radially-invariant quantities associated with the closed 2-forms ${*F}_\2$ and $F_\2$ after integrating out the angular coordinates.  Analogously, the NUT-induced electric and magnetic charges are then the angular-independent quantities after integrating radially from the horizon to infinity. We provided detail procedures of obtaining all the thermodynamic variables and we then verified that the first law was indeed satisfied. The results are also consistent with the Euclidean action and the Smarr relation.

The Taub-NUT and the Plebanski solutions are typically presented in literature in the coordinate system where the Misner strings are symmetrically located at the north and south poles. One can choose a coordinate gauge such that the Misner string exists only at the north or the south pole. We considered the more general coordinate gauge \eqref{asymt} with a free dimensionless parameter $\alpha$. Our well-defined procedures allowed us to obtain straightforwardly all the thermodynamic variables in this more complicated case. Although many thermodynamic variables such as the mass, temperature and entropy, etc.~depend nontrivially on $\alpha$, both $\alpha$ and $\delta \alpha$ drop out completely from the first law, indicating that $\alpha$ is nontrivial thermodynamically, splitting the degeneracy of the NUT charges of Misners strings at north and south poles. We also confirm that the free energy for this more general case is free from $\alpha$. This provides a strong validation of our approach since the Euclidean action should not change under \eqref{asymt}.

The mass we obtained from the generalized Komar form is no longer simply the parameter $m$, but modified by the NUT parameter. This is not surprising physically since the NUT charge creates the Misner strings that distort the spacetime metric that affects universally all matter. This however cannot be said about the Dirac strings that only affect matter that couples to the Maxwell field; therefore, the mass of spacetime does not depend on the Dirac strings. This leads to our different treatments of the contribution to the mass from the Dirac and Misner strings even though they have the analogous structure in the (generalized) Komar form. It is worth mentioning also that the elegant symmetry in our mass formula \eqref{nutmass3} for the Taub-NUT metric suggests that the parameters $(m,n)$ may be viewed as ``gravitational electric and magnetic'' contributions to the mass. Our findings reveal very rich and enormously interesting structures of Taub-NUT geometries, and may pave the way to decipher more general and complicated nutty spaces.

\section*{Acknowledgement}

We are grateful to Yi Pang and Ze Li for useful discussions, and to Shuang-Qing Wu for pointing out a few typos in the earlier versions. This work is supported in part by NSFC (National Natural Science Foundation of China) Grants No.~12075166, No.~11675144, No.~11875200 and No.~11935009.

\section*{Appendix}

\appendix

\section{Wald formalism and Generalized Komar integration}
\label{app:wald}

\subsection{Wald formalism}

\begin{figure}[hbtp]
  \centering
  \includegraphics[width=0.24\textwidth]{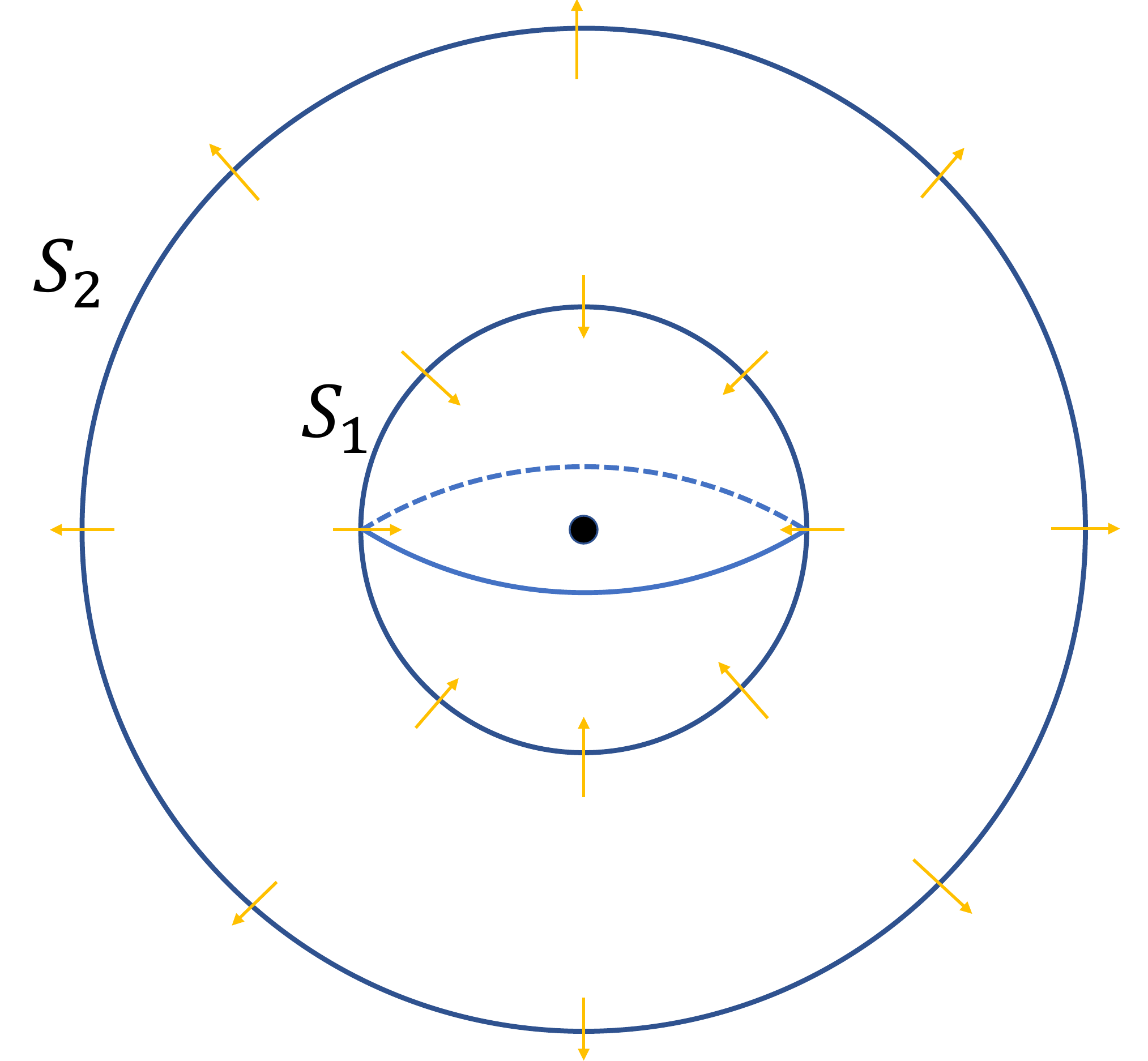}\ \
  \includegraphics[width=0.24\textwidth]{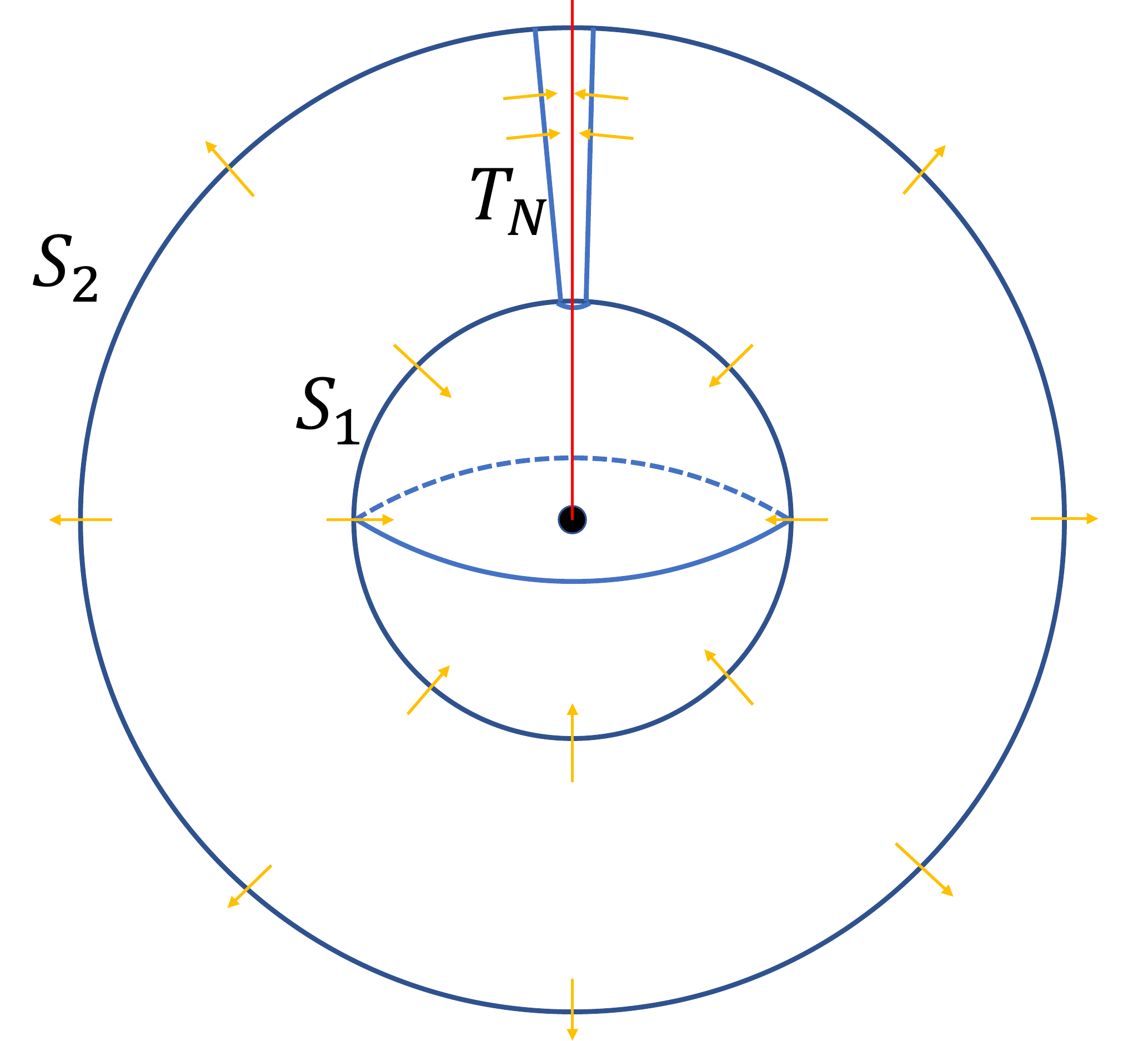}\
  \includegraphics[width=0.24\textwidth]{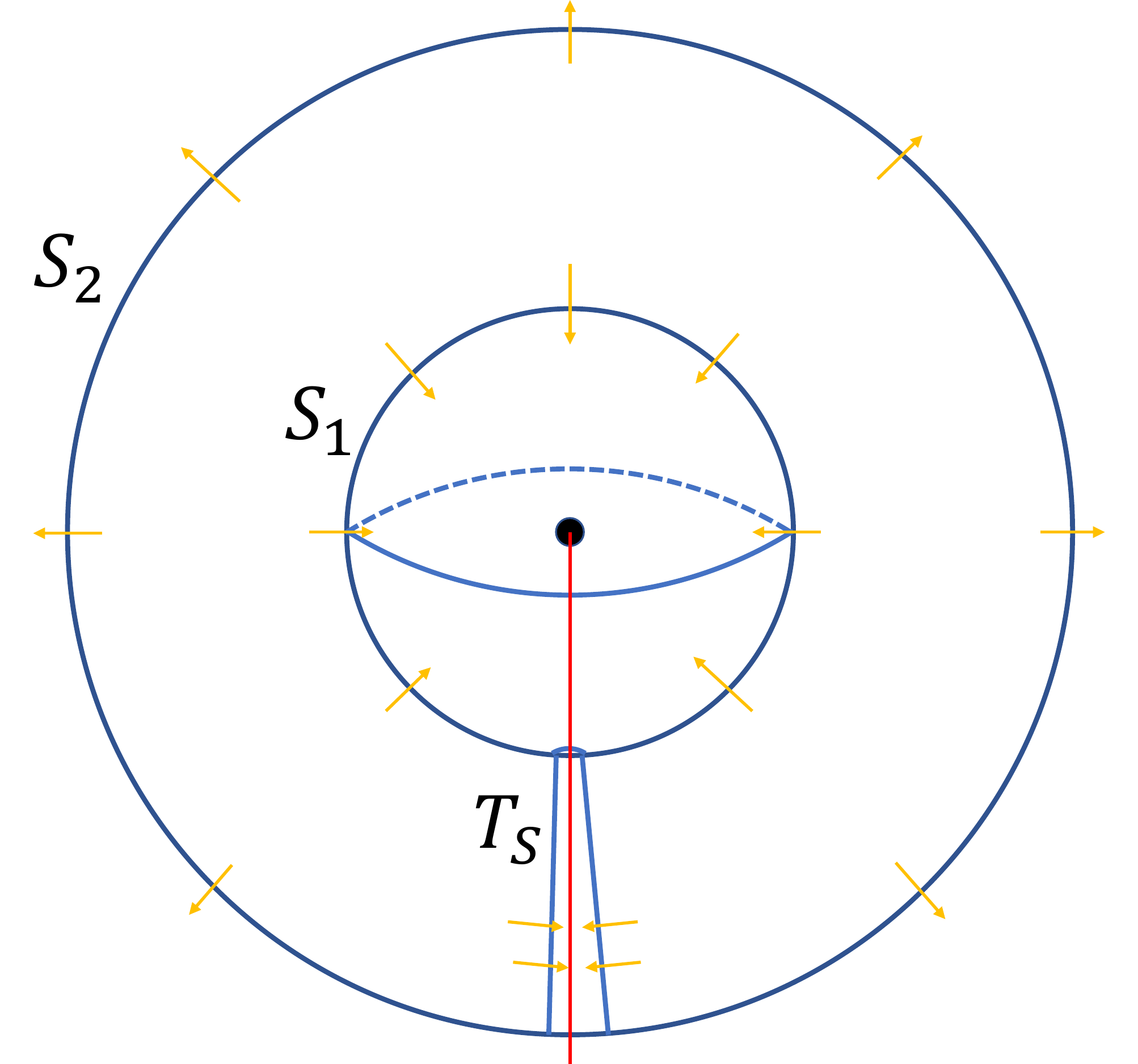}\ \
  \includegraphics[width=0.24\textwidth]{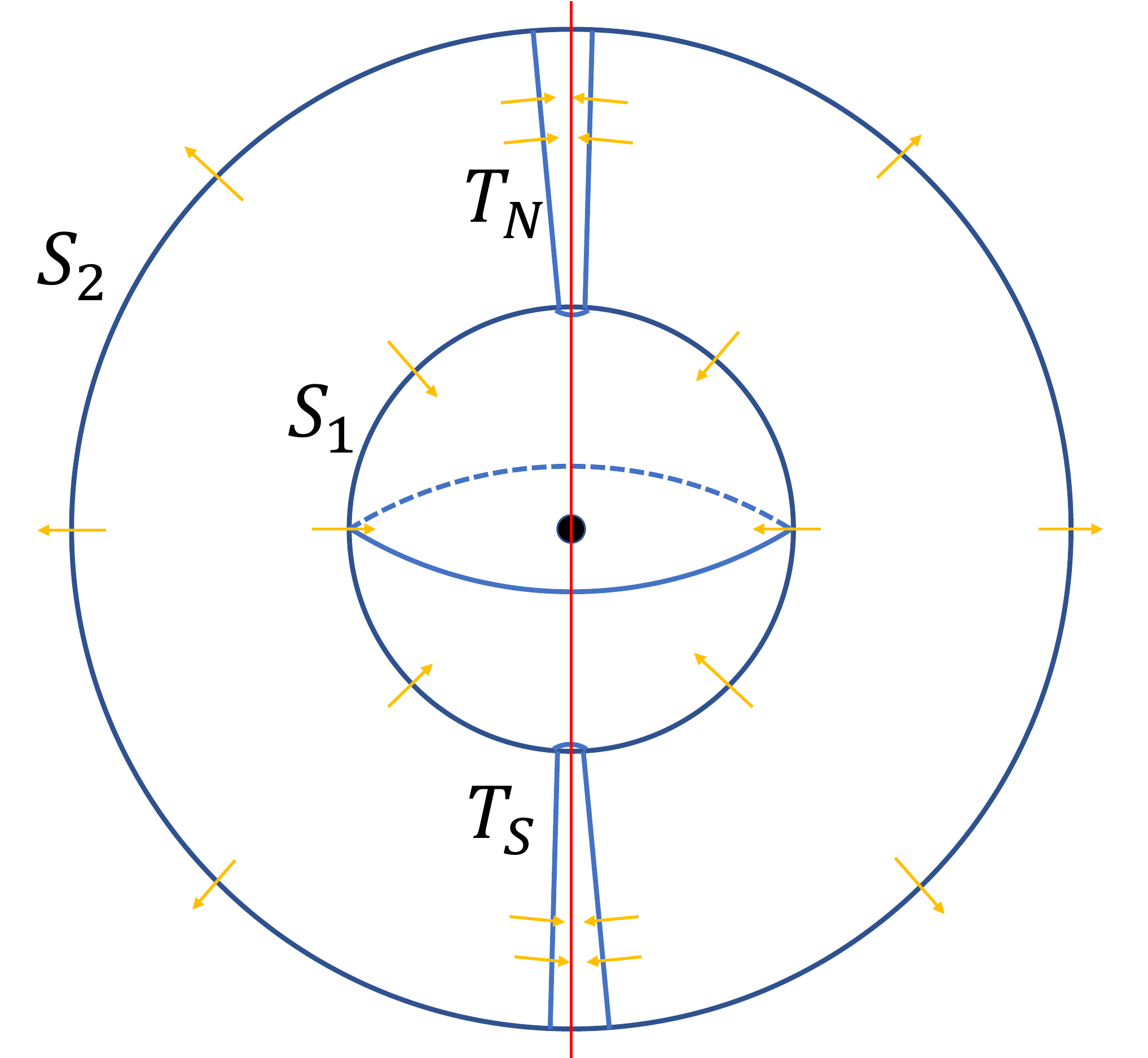}
 \caption{\small Codimension-2 hypersurfaces that surrounds the bulk, with or without various Misners or Dirac strings.}\label{plots}
\end{figure}

In this appendix, we describe the Wald approach to black hole dynamics for the general EMD theory \eqref{emdlag}, which we employ extensively in this paper. We shall only describe the key steps to obtain the infinitesimal Hamiltonian without any derivation. Those who are interested in the proof can read the original papers \cite{Wald:1993nt,Iyer:1994ys}. The EMD theory contains the metric $g_{\mu\nu}$, the Maxwell field $A_\mu$ and the dilaton scalar $\varphi$. The Wald formalism begins with the full variation
\be
\frac{\delta \mathcal{L}}{\sqrt{-g}}= E^g_{\mu\nu}\delta g^{\mu\nu}+E_A^{\mu}\delta A_{\mu}+E_\varphi\delta\varphi+\nabla_\mu\Theta^\mu\,,
\ee
with
\bea
E^g_{\mu\nu}&=& R_{\mu\nu}-\frac{1}{2}g_{\mu\nu}\frac{ \mathcal{L}}{\sqrt{-g}}-\frac{1}{2}(\nabla_\mu\varphi)(\nabla_\nu\varphi)-
\frac{1}{2}e^{a\varphi}F^2_{\mu\nu}\,,\nn\\
E_A^{\nu}&= &\nabla_\mu\left(e^{a\varphi}F^{\mu\nu}\right)\,,\qquad
E_\varphi=\Box\varphi-\frac{a}{4}e^{a\varphi}F^2,\qquad \Theta^\mu=\Theta_g^\mu+\Theta_A^\mu+\Theta_\varphi^\mu\,,\nn\\
\Theta_g^\mu&=& g^{\mu\alpha}\nabla^\beta\delta g_{\alpha\beta}-g^{\alpha\beta}\nabla^\mu\delta g_{\alpha\beta},\qquad \Theta_A^\mu=-e^{a\varphi}F^{\mu\nu}\delta A_{\nu}\,,\qquad \Theta_\varphi^\mu=-(\nabla^\mu\varphi)\delta\varphi\,.
\eea
The Noether current associated with a Killing vector $\xi$ takes the form
\be
J^\mu=\Theta^\mu-\xi^\mu\frac{\mathcal{L}}{\sqrt{-g}}-2E_g^{\mu\nu}\xi_\nu+
E_A^{\mu}A_{\lambda}\xi^\lambda\,,
\ee
and the corresponding Noether charge, defined by $J^\mu=\nabla_\nu Q^{\mu\nu}$, is
\be
Q^{\mu\nu}=-2\nabla^{[\mu}\xi^{\nu]}-e^{a\varphi}F^{\mu\nu}A_{\lambda}\xi^\lambda\,.\label{waldQ}
\ee
To proceed, one may define two $(D-2)$-forms
\bea
\mathbf{Q}[\xi]&=&\frac{1}{2!(D-2)!}\epsilon_{\alpha\beta\mu_1\mu_2\cdots\mu_{D-2}}
Q^{\alpha\beta}dx^{\mu_1}\wedge dx^{\mu_2}\cdots\wedge dx^{\mu_{D-2}},\cr
i_\xi\mathbf{\Theta} &=& \frac{1}{(D-2)!}\epsilon_{\alpha\beta\mu_1\mu_2\cdots\mu_{D-2}}
\Theta^{\alpha}\xi^{\beta}dx^{\mu_1}\wedge dx^{\mu_2}\cdots\wedge dx^{\mu_{D-2}}.
\eea
It was shown that the combination $(\delta {\mathbf Q}[\xi] - i_\xi {\mathbf \Theta})$ is closed, namely
\be
d(\delta {\mathbf Q}[\xi] - i_\xi {\mathbf \Theta})=0\,.
\ee
We can thus apply the Stokes theorem and obtain
\be
0=\delta {\cal H} =\fft{1}{16\pi} \int_{\rm bulk} (d(\delta {\mathbf Q}[\xi] - i_\xi {\mathbf \Theta})= \fft{1}{16\pi} \int_{\Sigma} (\delta {\mathbf Q}[\xi] - i_\xi {\mathbf \Theta})\,,\label{waldgen}
\ee
where $\Sigma$ is the codimension-two hypersurfaces that surround the bulk. In many black holes such as the Schwarzschild black hole or the electric RN black hole, the spacetime on and outside of the horizon is well defined. The hypersurfaces that surround the bulk is the Cauchy surface $S_1$ on the horizon and the foliating sphere $S_2$ at the asymptotic infinity, as shown in the first graph of Fig.~\ref{plots}.  We therefore have
\be
\Sigma = \Sigma_+(S_1) \bigcup \Sigma_\infty(S_2)\,.
\ee
It follows from \eqref{waldgen} that the first law of thermodynamics of these black holes is then the consequence of the identity
\be
\delta {\cal H}_\infty = \delta {\cal H}_+\,.\label{waldidentity1}
\ee

The situation becomes more complicated when Dirac or Misner strings exist in the bulk. Such strings can exist at the north pole, or the south pole or both.  We need to cutout these string singularities by using infinitesimal encompassing tubes at the north or south poles $T_{N}$ and $T_{S}$. The Wald identity based on the Stokes theorem now becomes
\bea
\Sigma = \Sigma_+(S_1) \bigcup \Sigma_\infty(S^2)\bigcup T_N:&&\qquad \delta{\cal H}_{S_2}+\delta{\cal H}_{S_1}+\delta{\cal H}_{T_N}=0\,,\cr
\Sigma = \Sigma_+(S_1) \bigcup \Sigma_\infty(S^2)\bigcup T_S:&&\qquad \delta{\cal H}_{S_2}+\delta{\cal H}_{S_1}+\delta{\cal H}_{T_S}=0\,,\cr
\Sigma = \Sigma_+(S_1) \bigcup \Sigma_\infty(S^2)\bigcup T_N\bigcup T_S:&&\qquad\delta{\cal H}_{S_2}+\delta{\cal H}_{S_1}+\delta{\cal H}_{T_N}+\delta{\cal H}_{T_S}=0\,.
\label{waldidentity2}
\eea
Such a classical identity underlies the corresponding black hole thermodynamic first law.

We now consider a concrete example, the four-dimensional electric black hole of the EMD theory studied in section 2. There are no Dirac or Misner strings in the bulk, and therefore the Wald identity \eqref{waldidentity1} holds. It is clear that the null Killing vector on the horizon is $\xi=\partial_t$. It is straightforward to obtain
\bea
&&{\mathbf Q}[\xi]=\left(r^2f'-\frac{Nr^2fH'}{2H}-r^2e^{a\varphi} H^{\frac{N}{2}}\psi_e'\psi_e\right)
\Omega_\2\,,\qquad \Omega_\2=\sin\theta d\theta\wedge d\varphi\,,\\
&&\delta {\mathbf Q}[\xi]-i_\xi {\mathbf \Theta}=\Big(-2r\delta f-\psi_e \delta\left(r^2e^{a\varphi}H^{\frac{N}{2}}\psi_e'\right) +\frac{Nr^2}{2H}\left(f'\delta H-H'\delta f\right)-\frac{N}{H}r^2f\delta H'
\Big)\Omega_\2\,.\nn
\eea
Here a prime denotes a derivative with respect to coordinate $r$. Substituting the explicit $(f,H,\psi)$ functions into the above, we find that ${\mathbf Q}[\xi]$ is not closed, but the 2-form  ($\delta {\mathbf Q}[\xi]-i_\xi {\mathbf \Theta}$) is indeed closed.  Specifically, we have
\be
\delta {\mathbf Q}[\xi]-i_\xi {\mathbf \Theta}=4\delta M\,\Omega_\2\,,\qquad \rightarrow\qquad
\delta {\cal H}=\fft{1}{16\pi} \int =\delta M\,,\label{emdelectricdeltah1}
\ee
where $M=\fft12(\mu + \fft12 N q)$, as given in \eqref{MTSforEMD}. (Note that $\int \Omega_\2=4\pi$.)
Since this is radially conserved, and hence it must also equal to the one given on the horizon:
\be
\delta {\cal H}_+ = T \delta S + \Phi_e \delta Q_e\,.\label{emdelectricdeltah2}
\ee
Note that we made use of \eqref{varychi} to evaluate $\delta {\cal H}_+$ on the horizon.
We also made use of the fact that $r^2e^{a\varphi}H^{\frac{N}{2}}\psi_e'=-4Q_e$ is the first integral of the Maxwell equation. Combining \eqref{emdelectricdeltah1} and \eqref{emdelectricdeltah2} leads to the first law. We discussed in section 2 the subtler magnetic case where Dirac strings are involved.

\subsection{Generalized Komar form}

The Komar integration is a purely geometric quantity, integrating over a Komar $(D-2)$-form
\be
\mathbf{Q}_{\rm K}[\xi]=\frac{1}{2!(D-2)!}\epsilon_{\alpha\beta\mu_1\mu_2\cdots\mu_{D-2}}
Q^{\alpha\beta}dx^{\mu_1}\wedge dx^{\mu_2}\cdots\wedge dx^{\mu_{D-2}}\,,\qquad
Q^{\mu\nu}_{\rm K}=-2\nabla^{[\mu}\xi^{\nu]}\,.
\ee
For simplicity, we shall focus on $D=4$ dimensions. For the simple spacetimes illustrated as the first graph in Fig.~\ref{plots}, evaluating the Komar integration over $S_2$ at infinity give the Komar mass $M$ with an appropriate overall coefficient, for $\xi=\partial_t$, i.e.
\be
M = \fft{1}{8\pi} \int_{S_2} \mathbf{Q}_{\rm K}\,.
\ee
Evaluating on the horizon gives $2TS$. The strong energy condition of the matter sector ensures that $M\ge 2 TS$, saturated by the Schwarzschild black hole. It was recently shown that the inequality can also be independently satisfied by the combination of the null and trace energy conditions \cite{Khodabakhshi:2022jot}. In pure gravity, the Komar form field is closed, i.e.~$d(\mathbf{Q}_{\rm K}[\xi])=0$, giving rise to the Smarr relation $M=2TS$ for the Schwarzschild black hole. In this case, the spacetime does not have Misner string singularities, and integrating over the latitude angle gives a radially-invariant quantity $M$. When string singularities are involved, then we need cutout the infinitesimal tubes that encompass the strings, as shown in Fig.~\ref{plots}. In either case, $\int_{\Sigma} \mathbf Q_{\rm K}=0$ gives the Smarr relation.

In pure gravity, the ${\mathbf Q}[\xi]$ in the Wald formalism is identical to the Komar form. The Wald formalism illustrates that the Komar mass contributes only half to the infinitesimal Hamiltonian, since it enters the Wald formalism with with the factor ``$\fft{1}{16\pi}$'' rather than ``$\fft{1}{8\pi}$.''  The Wald $\mathbf Q[\xi]$ can be viewed as some generalization of the Komar form, when matter is involved. However, it is not closed. Here we would like to present a generalization of the Komar form that is also closed. For the EMD theory, it has an extra term compared to \eqref{waldQ} in the Wald formalism. We have
\be
\widetilde Q^{\mu\nu}=-2\nabla^{[\mu}\xi^{\nu]}-e^{a\varphi}\,F^{\mu\nu}A_{\lambda}\xi^\lambda
+ e^{a\varphi}\,A_\rho F^{[\mu |\rho|} \xi^{\nu]}\,.
\ee
In the form language, we have
\be
\widetilde {\mathbf Q}[\xi]=-{*d}\xi-\ft{1}{2}e^{a\varphi}{*F}_{\2}\left(i_\xi A_{\1}\right)+\ft{1}{2}e^{a\varphi}\left(i_\xi {*F}_{\2}\right)\wedge A_\1\,.\label{emdtildeQ}
\ee
For the electrically charged black hole, it is given by
\be
\widetilde {\mathbf Q}[\xi]=\Big(r^2 f' - \fft{N r^2 f H'}{2H} - \ft12 r^2 e^{a\varphi} H^{\fft12 N} \psi \psi'\Big)\Omega_\2= 2 M \Omega_\2\,,
\ee
where $M$ was given under \eqref{emdelectricdeltah1}. The second equality above shows manifestly that $d\widetilde {\mathbf Q}[\xi]=0$. Evaluate the middle term on the horizon, we obtain the Smarr relation
\be
M=2 T S + \Phi_e Q_e\,.
\ee
The situation for the magnetic solution is rather different, and we discussed this in the main text in section 2.

\end{document}